\documentclass[12pt,a4paper]{article}
\usepackage{amssymb}
\font\oneeight=cmr10 at 18pt

\newcommand{\vTm}{\vphantom{\mbox{\oneeight I}}}

\makeatletter
\def\eqnarray{%
\stepcounter{equation}%
\let\@currentlabel=\theequation
\global\@eqnswtrue
\global\@eqcnt\z@
\tabskip\@centering
\let\\=\@eqncr
$$\halign to \displaywidth\bgroup\@eqnsel\hskip\@centering
$\displaystyle\tabskip\z@{##}$&\global\@eqcnt\@ne
\hfil$\displaystyle{{}##{}}$\hfil
&\global\@eqcnt\tw@$\displaystyle\tabskip\z@{##}$\hfil
\tabskip\@centering&\llap{##}\tabskip\z@\cr}
\makeatother
\newcommand{\kansu}[2]{{{#1}\!\left({#2}\right)}}
\newcommand{\ket}[1]{{\vert{#1}\rangle}}
\newcommand{\bra}[1]{{\langle{#1}\vert}}
\newcommand{\kett}[2]{{\vert{#1,#2}\rangle}}
\newcommand{\ai}{{\sqrt{-1}}}
\newcommand{\calh}{{\cal H}}
\newcommand{\calm}{{\cal M}}
\newcommand{\cala}{{\cal A}}
\newcommand{\calf}{{\cal F}}

\newcommand{\integer}{{\mathbf Z}}
\newcommand{\fukuso}{{\mathbf C}}
\newcommand{\real}{{\mathbf R}}
\newcommand{\futon}{{\bf N}}

\newcommand{\stm}{{St_m}}
\newcommand{\grm}{{Gr_m}}
\newcommand{\eem}{{E_m}}

\newcommand{\lam}{{\bf \lambda}}
\newcommand{\slam}{{\bf \lambda_0}}

\begin{document}

\title{\sl Introduction to Grassmann Manifolds and Quantum Computation}
\author{
   Kazuyuki FUJII
   \thanks{E-mail address: fujii@yokohama-cu.ac.jp}\\
   Department of Mathematical Sciences\\
   Yokohama City University\\
   Yokohama 236-0027\\
   JAPAN
   }
\date{}
\maketitle\thispagestyle{empty}
\renewcommand{\theequation}{\arabic{section}.\arabic{equation}}
%
%
%
%
\begin{abstract}
   Geometrical aspects of
   quantum computing are reviewed elementarily for  non-experts and/or
   graduate students who
   are interested in both Geometry and Quantum Computation.

   In the first half we show how to treat Grassmann manifolds which
   are very important examples of manifolds in Mathematics and Physics.
   Some of their applications  to Quantum Computation and its  efficiency
   problems are shown in the second half.
   An interesting current topic of Holonomic Quantum Computation is also
   covered.

   In the Appendix  some related advanced topics are discussed.
\end{abstract}


%
%
%
%

\section{Introduction}
\label{intro}
\setcounter{equation}{0}

This is a review article based on lectures given at several universities in
Japan and a talk at Numazu-meeting\footnote{A meeting held
by Yoshinori Machida at Numazu College of Technology to discuss recent results
on Geometry, Mathematical Physics, String Theory, Quantum Computation, etc.}.
The aim is to show a  somewhat unconventional  but fruitful path connecting
Geometry and Quantum
Computation, and the audience is graduate students and/or non-experts
who are interested in both of the disciplines.

The progress of Quantum Computation after the excellent work of P. Shor
\cite{PS} on prime factorization of integers  and the work of L. Grover
\cite{LG} on quantum data-base searching is very remarkable.
These discoveries have given  great impacts on scientists. They drove not only
theoreticians to finding other quantum algorithms, but also
experimentalists to building practical quantum computers.
For standard introduction see for example \cite{AS}, \cite{RP}, \cite{LPS} or
\cite{AH}.

The conventional methods of Quantum Computation are  more or less
  algebraic.
On the other hand we are interested in geometrical or topological methods.
Geometry or Topology are crucial to understand mathematical or physical
objects from the global point of view.

For general introduction of Geometry and Topology the book \cite{MN} is
strongly recommended. But in this book the volume calculations of some
important manifolds like Grassmann ones or more generally symmetric spaces
are missing. They are important in understanding of entanglements or
entangled measures.
In the first half we show in some detail the volume calculations of Grassmann
manifolds.
Here let us recall some basic concepts.

A homogeneous space is defined by
\[
     M \cong G/H,
\]
where $G$ is a Lie group and $H$ its subgroup. We are particularly
interested in the case where $G$ is a classical group (for example,
a unitary group $U(n)$ or an orthogonal group $O(n)$). The complex Grassmann
manifold $G_{k,n}(\fukuso)$, which is our main concern in this paper, is 
written
as
\[
     G_{k,n}(\fukuso) \cong \frac{U(n)}{U(k)\times U(n-k)}.
\]
The volume of $G_{k,n}$ is expressed in terms of the well-known
volume of $U(n)$  (see the following sections),
\[
   \mbox{Vol}\left(G_{k,n}(\fukuso)\right) = \frac{\mbox{Vol}\left(U(n)\right)}
                             {\mbox{Vol}\left(U(k)\right)\times 
\mbox{Vol}\left(U(n-k)\right)}.
\]
This is the usual method to obtain the volume of homogeneous spaces.

On the other hand, the volume is obtained by integrating
  the volume form of Grassmann manifolds (:
$dv(Z,Z^{\dagger})$) which is  expressed in terms of local
coordinates (: $Z$) (see the following sections):
\[
    \mbox{Vol}(G_{k,n}(\fukuso)) = \int_{G_{k,n}(\fukuso)}dv(Z,Z^{\dagger}).
\]
Is it really possible (practical) to carry out the
integral on the right hand side?  As far as we know such a calculation
has not been performed except for $k=1$ (case of complex projective spaces).
For $k \geq 2$ direct calculation seems to be very complicated.
We would like to present this calculation as a challenging problem to
the readers.

\vspace{3mm}
Let us come back to Quantum Computation (QC briefly).

\par \noindent
Gauge theories are widely recognized as the basic ingredients
of quantum field theories which enjoy
remarkable progress recently, String Theory, M-Theory,
F-Theory, etc.
Therefore it is very natural to incorporate gauge theoretical ideas
to QC;  that is construction of ``{\bf gauge theoretical}'' quantum
computation and/or of ``{\bf geometric}'' quantum computation in our 
terminology.
The merit of geometric (or topological) method of QC may be the stability
with respect to the
influence from the environment.

In \cite{ZR} and \cite{PZR} Zanardi and Rasetti proposed an attractive idea
$\cdots$ {\bf Holonomic Quantum Computation} $\cdots$ using the non-abelian
Berry phase (quantum holonomy in the mathematical terminology).
We introduce this concept in the final  section.
See also \cite{AYK} and \cite{JP} for another interesting geometric model.

Quantum Computation comprises of many subjects.
To give a comprehensive overview
is beyond the scope of this article,
so we focus our attention on the construction and
the efficiency of unitary operations, and give  geometric interpretation to 
them.
Here let us make a brief review.

For $n=2^{t}\ (t \in \futon)$ we set a unitary operation
\[
U_{f} : (\fukuso^{2})^{\otimes t} \longrightarrow (\fukuso^{2})^{\otimes t}\
;\  U_{f}(|a))=(-1)^{f(a)}|a)
\]
where $f$ is a signature function defined by
\[
f : \{0,1,\cdots,n-1\} \longrightarrow \integer_{2}=\{0,1\},
\quad a \mapsto f(a)
\]
and
\begin{eqnarray}
&&|a)\equiv
\ket{a_1}\otimes \ket{a_2}\otimes \cdots \ket{a_{t-1}}\otimes \ket{a_t},
\quad a_{k} \in \integer_{2}
\nonumber \\
&&a=a_{1}2^{t-1}+a_{2}2^{t-2}+\cdots+a_{t-1}2+a_t ,\quad  0 \leq a \leq n-1 .
\nonumber
\end{eqnarray}
This operation plays a crucial role in the quantum data-base searching
algorithm of Grover, \cite{LG} and an important role in quantum computing, 
in general.
Our concern is as follows. Is it possible to construct this operator
in an efficient manner (steps polynomial in $t$)?  Has such an algorithm
been already given in Quantum Computation?

\par \indent
As far as we know this point is rather unclear, \cite{KFu}.
See \cite{9th} and \cite{KF6}.  We will discuss this point in some detail .

We would like to construct a road connecting Geometry and Quantum Computation,
which is not an easy task. We will show one of such attempts as explicitly
as possible.
Though  results given in this paper are not new,
we do hope our presentation offers new perspectives
to not only students and/or non-experts but also to experts.

{\bf The contents of this paper are as follows}:
\begin{itemize}
\item[1] Introduction
\item[2] Grassmann Manifolds
\item[3] Volume of Unitary Groups
\item[4] A Question
\item[5] Quantum Computing
\item[6] Holonomic Quantum Computation
\item[] Appendix
  \begin{itemize}
  \item[A] A Family of Flag Manifolds
  \item[B] A Generalization of Pauli Matrices
  \item[C] General Controlled Unitary Operations
  \end{itemize}
\end{itemize}

\section{Grassmann Manifolds}
\label{manifold}
\setcounter{equation}{0}

Let $V$ be a $k$-dimensional subspace in $\fukuso^n$ ($0\leq k \leq n$).
Then it is well-known in Linear Algebra that there is only one projection
$P : \fukuso^n \longrightarrow \fukuso^n$ with $V=P(\fukuso^n)$. Here
the projection means $P^2=P$ and $P^{\dagger}=P$ in $M(n;\fukuso)$.

\par \noindent
The Grassmann manifold is in this case defined by all the $k$-dimensional 
subspaces
in $\fukuso^n$, and it is identified with all the projections in $M(n;\fukuso)$
with the trace $k$ or the rank $k$ (corresponding to $V=P(\fukuso^n)$). We 
note that
the eigenvalues of a projection are either $0$ or $1$ (by $P^2=P$),
so the rank of $P$ =
trace of $P$. Therefore we arrive at
\begin{equation}
   \label{eq:grassmann-definition}
   G_{k,n}(\fukuso)=\{P \in M(n;\fukuso) |\ P^2=P,\ P^{\dagger}=P\
       \mbox{and}\ \mbox{tr}P=k \}.
\end{equation}
A comment is in order. In general it is not easy to visualize all the 
$k$-dimensional
subspaces in $\fukuso^n$ except for experts in Geometry.
But it is easy even for us to deal with (\ref{eq:grassmann-definition})
as will be shown in the following.

\par \noindent
We note that $G_{0,n}(\fukuso)=\{{\bf 0}_n \}$ and $G_{n,n}(\fukuso)=
\{{\bf 1}_n \}$.
In particular $G_{1,n}(\fukuso)$ is called a complex projective space and is
written as ${{\fukuso}P}^{n-1}$. In (\ref{eq:grassmann-definition}) we know
a natural symmetry (isomorphism)
\begin{equation}
     \kappa : G_{k,n}(\fukuso) \longrightarrow G_{n-k,n}(\fukuso),\quad
     \kappa (P)={\bf 1}_{n}- P ,
\end{equation}
so that we have $G_{k,n}(\fukuso) \cong G_{n-k,n}(\fukuso)$.

Now it is easy to see that $P$ can be written as
\begin{equation}
   \label{eq:diagonal}
      P=AE_{k}A^{-1}\quad \mbox{for some}\  A \in U(n),
\end{equation}
where $E_{k}$ is a special projection
\begin{equation}
\label{eq:special-projection}
E_{k}=
   \left(
     \begin{array}{cc}
        {\bf 1}_{k}& O     \\
         O&    {\bf 0}_{n-k}
     \end{array}
   \right).
\end{equation}
Therefore we have
\begin{equation}
    \label{eq:grassmann-henkei}
   G_{k,n}(\fukuso)=\{AE_{k}A^{-1} |\ A \in U(n) \},
\end{equation}
which directly leads  to
\begin{equation}
     G_{k,n}(\fukuso) \cong \frac{U(n)}{U(k)\times U(n-k)}\ .
\end{equation}
In particular
\begin{equation}
     G_{1,n}(\fukuso) = {{\fukuso}P}^{n-1}
                        \cong \frac{U(n)}{U(1)\times U(n-1)}\
                        \cong \frac{U(n)/U(n-1)}{U(1)}\
                        \cong \frac{\mbox{S}^{2n-1}}{\mbox{S}^{1}}\ ,
\end{equation}
see (\ref{eq:odd-sphere}). Here $\mbox{S}^{k}$ is the unit sphere in
${\real}^{k+1}$ and $U(1)$ = $\mbox{S}^{1}$.\
We note that $G_{k,n}(\fukuso)$ is a complex manifold (moreover, a
K\"ahler manifold) and its complex dimension is $k(n-k)$.

Next let us introduce  local coordinates around $P$ in (\ref{eq:diagonal}).
We denote by  $M(n-k,k;\fukuso)$  the set of all $(n-k)\times k$ - matrices 
over
$\fukuso$ and define a map
\[
   {\cal P} : M(n-k,k;\fukuso) \longrightarrow G_{k,n}(\fukuso)
\]
as follows :
\begin{equation}
    \label{eq:local-coordinate}
    {\cal P}(Z) = A
           \left(
             \begin{array}{cc}
                  {\bf 1}_{k}& - Z^{\dagger} \\
                      Z&    {\bf 1}_{n-k}
             \end{array}
           \right)
           \left(
             \begin{array}{cc}
                  {\bf 1}_{k}&  O \\
                      O&    {\bf 0}_{n-k}
             \end{array}
           \right)
           \left(
             \begin{array}{cc}
                  {\bf 1}_{k}& - Z^{\dagger} \\
                      Z&    {\bf 1}_{n-k}
             \end{array}
           \right)^{-1}
           A^{-1}.
\end{equation}
Of course ${\cal P}({\bf 0})$ = $P$ in (\ref{eq:diagonal}).
\par
\noindent
Here a natural question arises. How many local coordinates do we have on
$G_{k,n}(\fukuso)$ ?  The number of them is just $_nC_k$.

\par \noindent
A comment is in order. We believe that this is the best choice of
local coordinates on the Grassmann manifold, and this one is called
the Oike coordinates in Japan.
As far as the author knows  H. Oike is the first to write down
(\ref{eq:local-coordinate}), \cite{HO}.
\par
\noindent  From this we can show the curvature form
${\cal P}(Z)d{\cal P}(Z)\wedge d{\cal P}(Z)$:
\begin{equation}
   \label{eq:bibun}
    d{\cal P}(Z) = A
           \left(
             \begin{array}{cc}
                  {\bf 1}_{k}& - Z^{\dagger} \\
                      Z&    {\bf 1}_{n-k}
             \end{array}
           \right)
           \left(
             \begin{array}{cc}
                  {\bf 0}_{k}& {\Lambda_{k}}^{-1}dZ^{\dagger}  \\
                     {M_{n-k}}^{-1}dZ& {\bf 0}_{n-k}
             \end{array}
           \right)
           \left(
             \begin{array}{cc}
                  {\bf 1}_{k}& - Z^{\dagger} \\
                      Z&    {\bf 1}_{n-k}
             \end{array}
           \right)^{-1}
           A^{-1},
\end{equation}
\begin{eqnarray}
   \label{eq:curvature}
    &&{\cal P}(Z)d{\cal P}(Z)\wedge d{\cal P}(Z) \nonumber \\
    &&= A
           \left(
             \begin{array}{cc}
                  {\bf 1}_{k}& - Z^{\dagger} \\
                      Z&    {\bf 1}_{n-k}
             \end{array}
           \right)
           \left(
             \begin{array}{cc}
                 {\Lambda_k}^{-1}dZ^{\dagger}\wedge {M_{n-k}}^{-1}dZ & O  \\
                   O  & {\bf 0}_{n-k}
             \end{array}
           \right)
           \left(
             \begin{array}{cc}
                  {\bf 1}_{k}& - Z^{\dagger} \\
                      Z&    {\bf 1}_{n-k}
             \end{array}
           \right)^{-1}
           A^{-1},
\end{eqnarray}
where
\begin{equation}
   \Lambda_{k} = {\bf 1}_{k}+Z^{\dagger}Z \in M(k;\fukuso), \quad
   M_{n-k} =  {\bf 1}_{n-k}+ZZ^{\dagger} \in M(n-k;\fukuso).
\end{equation}
In the following we omit the $\wedge$ symbol and write, for example,
${\cal P}d{\cal P}d{\cal P}$ instead of
${\cal P}(Z)d{\cal P}(Z)\wedge d{\cal P}(Z)$ for simplicity.
A (global) symplectic 2-form on $G_{k,n}(\fukuso)$ is given by
\[
        \omega = \mbox{tr}{\cal P}d{\cal P}d{\cal P}
\]
and its local form
\begin{equation}
   \label{eq:omega}
    \omega
   = \mbox{tr}\left( \Lambda_k^{-1}dZ^{\dagger}M_{n-k}^{-1}dZ \right)
           = \mbox{tr}\left(\vTm ({\bf 1}_{k}+Z^{\dagger}Z)^{-1}dZ^{\dagger}
                   ({\bf 1}_{n-k}+ZZ^{\dagger})^{-1}dZ \right).
\end{equation}

We want to rewrite (\ref{eq:omega}). Before doing this
let us make some mathematical preliminaries. For $A \in M(m,\fukuso)$ and
$B \in M(n,\fukuso)$ a tensor product $A\otimes B$ of $A$ and $B$ is
defined as
\[
     A\otimes B = \left( a_{ij}B \right)\quad  \mbox{for}\ A=(a_{ij})\
        \mbox{and} \ B=(b_{pq}).
\]
For example,  for
\[
   A =
           \left(
             \begin{array}{cc}
                  a_{11} & a_{12} \\
                  a_{21}&  a_{22}
             \end{array}
            \right)
\quad \mbox{and}\quad
   B =
           \left(
             \begin{array}{cc}
                  b_{11} & b_{12} \\
                  b_{21}&  b_{22}
             \end{array}
            \right)
\]
we have
\begin{equation}
    \label{eq:fourbyfour}
    A\otimes B =
           \left(
             \begin{array}{cc}
                  a_{11}B & a_{12}B \\
                  a_{21}B & a_{22}B
             \end{array}
            \right)
         =
           \left(
             \begin{array}{cccc}
                a_{11}b_{11}& a_{11}b_{12}& a_{12}b_{11}& a_{12}b_{12} \\
                a_{11}b_{21}& a_{11}b_{22}& a_{12}b_{21}& a_{12}b_{22} \\
                a_{21}b_{11}& a_{21}b_{12}& a_{22}b_{11}& a_{22}b_{12} \\
                a_{21}b_{21}& a_{21}b_{22}& a_{22}b_{21}& a_{22}b_{22}
             \end{array}
            \right).
\end{equation}
Therefore  componentwise we have $(A\otimes B)_{ip,jq}=A_{ij}B_{pq}$.
Then it is not difficult to see
\begin{equation}
   \label{eq:tr-det}
    \mbox{tr}(A\otimes B) = \mbox{tr}(A)\mbox{tr}(B), \quad
    \mbox{det}(A\otimes B) = \{\mbox{det}(A) \}^{n}\{\mbox{det}(B) \}^{m}.
\end{equation}

Let us construct a column vector ${\widehat Z}$ in $\fukuso^{k(n-k)}$ from 
$Z$ in
$M(n-k,k;\fukuso)$ in a usual manner:
\[
  {\widehat Z} = \left(z_{11}, \cdots,z_{1k}, \cdots  \cdots,
          z_{{n-k},1}, \cdots, z_{{n-k},k} \right)^{T}\ ,
\]
where $T$ means a transpose.\  Now we rewrite (\ref{eq:omega}) as follows:
\begin{eqnarray}
      \label{eq:omega-2}
    \omega &=& \mbox{tr}\left( \Lambda_k^{-1}dZ^{\dagger}
                               M_{n-k}^{-1}dZ \right)
            = \mbox{tr}\left( dZ^{\dagger}M_{n-k}^{-1}
                               dZ\Lambda_k^{-1} \right)   \nonumber \\
           &=& \sum (dZ^{\dagger})_{ij}(M_{n-k}^{-1})_{jp}(dZ)_{pq}
                    (\Lambda_k^{-1})_{qi}
            = \sum d{\bar z}_{ji}(M_{n-k}^{-1})_{jp}dz_{pq}
                    (\Lambda_k^{-1})_{qi}                 \nonumber \\
           &=& \sum d{\bar z}_{ji}(M_{n-k}^{-1})_{jp}
                    (\Lambda_k^{-1})_{qi}dz_{pq}
            = \sum d{\bar z}_{ji}(M_{n-k}^{-1})_{jp}
                    {{(\Lambda_k^{-1})}^{T}}_{iq}dz_{pq}    \nonumber \\
           &=& \sum ({d{\widehat Z}^{\dagger})}_{ji} \left\{ M_{n-k}^{-1}
               \otimes (\Lambda_k^{-1})^{T} \right\}_{ji,pq}
                  d{\widehat Z}_{pq}
            = (d{\widehat Z})^{\dagger}\left\{ M_{n-k}^{-1}\otimes
                     (\Lambda_k^{-1})^{T} \right\} d{\widehat Z}.
\end{eqnarray}
The symplectic volume on $G_{k,n}(\fukuso)$ which coincides with the
usual volume is given by
\begin{equation}
   \label{eq:symplectic-volume}
   dv = \frac{1}{ \{k(n-k)\}! } \left(\frac{\omega}{2\ai}\right)^{k(n-k)}.
\end{equation}
Here $\frac{1}{2\ai}$ is a normalization factor.  From (\ref{eq:omega-2})
it is easy to see
\begin{equation}
   \omega^{k(n-k)} = \{k(n-k)\}!\  \mbox{det}\left\{ M_{n-k}^{-1}\otimes
         (\Lambda_k^{-1})^{T} \right\} \prod_{i,j} d{\bar z}_{ij}dz_{ij}.
\end{equation}
Therefore (\ref{eq:symplectic-volume}) becomes
\begin{equation}
   \label{eq:symplectic-volume-2}
      dv =  \mbox{det}\left\{ M_{n-k}^{-1}\otimes
         (\Lambda_k^{-1})^{T} \right\} \prod_{i,j}
         \frac{d{\bar z}_{ij}dz_{ij}}{2\ai}.
\end{equation}
On the other hand by (\ref{eq:tr-det}) we have
\begin{equation}
  \label{eq:Lambda-M}
    \mbox{det}\left\{ M_{n-k}^{-1}\otimes (\Lambda_k^{-1})^{T} \right\}
    =
      \left(\mbox{det}M_{n-k}^{-1} \right)^{k}
      \left(\mbox{det}(\Lambda_k^{-1}) \right)^{n-k}
    = \left(\mbox{det} M_{n-k} \right)^{-k}
      \left(\mbox{det} \Lambda_k \right)^{-(n-k)}.
\end{equation}
Here we note $\mbox{det}\Lambda_k$ = $\mbox{det}M_{n-k}$. \ For
\[
   X =
           \left(
             \begin{array}{cc}
                  {\bf 1}_{k}& - Z^{\dagger}\\
                  Z & {\bf 1}_{n-k}
             \end{array}
            \right),
\]
we have
\[
   \mbox{det}X
    = \mbox{det}
           \left(
             \begin{array}{cc}
                  {\bf 1}_{k}& - Z^{\dagger}\\
                  Z & {\bf 1}_{n-k}
             \end{array}
            \right)
    = \mbox{det}
           \left(
             \begin{array}{cc}
                  {\bf 1}_{k}+Z^{\dagger}Z & - Z^{\dagger}\\
                  O & {\bf 1}_{n-k}
             \end{array}
            \right)
    = \mbox{det}\left({\bf 1}_{k}+Z^{\dagger}Z\right)
    = \mbox{det}\Lambda_{k}.
\]
On the other hand
\[
   \mbox{det}X
    = \mbox{det}
           \left(
             \begin{array}{cc}
                  {\bf 1}_{k}& - Z^{\dagger}\\
                  Z & {\bf 1}_{n-k}
             \end{array}
            \right)
    = \mbox{det}
           \left(
             \begin{array}{cc}
                  {\bf 1}_{k} & - Z^{\dagger}\\
                  O & {\bf 1}_{n-k}+ZZ^{\dagger}
             \end{array}
            \right)
    = \mbox{det}\left({\bf 1}_{n-k}+ZZ^{\dagger}\right)
    = \mbox{det}M_{n-k},
\]
so that
\[
    \mbox{det}\Lambda_{k} = \mbox{det}M_{n-k} \quad \blacksquare
\]
       From (\ref{eq:Lambda-M})
$\mbox{det}\left\{ M_{n-k}^{-1}\otimes (\Lambda_k^{-1})^{T} \right\}$ =
$\left(\mbox{det}{\Lambda_k}\right)^{-n}$, so we arrive at
\begin{equation}
     dv(Z,Z^{\dagger}) = \left(\mbox{det}{\Lambda_k}\right)^{-n}
      \prod_{i,j} \frac{d{\bar z}_{ij}dz_{ij}}{2\ai} =
     \left\{\mbox{det}({\bf 1}_{k}+Z^{\dagger}Z)\right\}^{-n}
     \prod_{i,j} \frac{d{\bar z}_{ij}dz_{ij}}{2\ai}.
\end{equation}
         From the above-mentioned facts the volume of Grassmann manifold
$G_{k,n}(\fukuso)$ is given as
\begin{equation}
   \label{eq:grassmann-integral}
   \mbox{Vol}(G_{k,n}(\fukuso)) =
  \int_{M(n-k,k;\fukuso)} \frac{\prod_{i,j} \frac{d{\bar z}_{ij}dz_{ij}}{2\ai}
           }{ \left\{\mbox{det}({\bf 1}_{k}+Z^{\dagger}Z)\right\}^{n} }\ .
\end{equation}
\par
\begin{flushleft}
{\bf Problem}\quad  How can we calculate this integral ?
\end{flushleft}

\section{Volume of Unitary Groups}
\label{univolume}
\setcounter{equation}{0}
Here we will show a heuristic method of evaluation of the volume of unitary 
group $U(n)$.
  Let $S^{2k-1}$ be the  $2k-1$ - dimensional unit sphere ($k \geq 1$)
over $\real$ and the volume be $\mbox{Vol}(S^{2k-1})$. For example
$\mbox{Vol}(S^{1})$ = $2\pi$ and $\mbox{Vol}(S^{3})$ = $2{\pi}^2$.
In general we have
\begin{equation}
   \label{eq:sphere-volume}
   \mbox{Vol}(S^{2k-1}) = \frac{2{\pi}^k}{(k-1)!}.
\end{equation}
Since we know the fact
\begin{equation}
    \label{eq:odd-sphere}
       \frac{U(k)}{U(k-1)} \cong S^{2k-1},
\end{equation}
we have
\begin{eqnarray}
   \label{eq:fiber-bundle}
   U(n) &\doteq& \frac{U(n)}{U(n-1)}\times \frac{U(n-1)}{U(n-2)}\times \cdots
              \times  \frac{U(2)}{U(1)}\times U(1) \nonumber \\
    &\doteq& S^{2n-1}\times S^{2n-3}\times \cdots \times S^{3}\times S^{1},
\end{eqnarray}
where $\doteq$ means {\bf almost equal}!
\par
\noindent
A comment is in order. Of course the equality  does not
hold in (\ref{eq:fiber-bundle}) except
for the cases of $n = 1,\ 2$.
But for the purpose of volume-counting or cohomology-counting
there is no problem to use (\ref{eq:fiber-bundle})\footnote{You will know this
``questionable equation" may be rather useful, see for example \cite{RRS}.}.

\par
\noindent
By combining (\ref{eq:fiber-bundle}) and (\ref{eq:sphere-volume}) we
obtain
\begin{equation}
  \label{eq:unitary-volume}
  \mbox{Vol}(U(n)) = \prod_{j=1}^{n} \mbox{Vol}(S^{2j-1})
                   = \prod_{j=1}^{n} \frac{2{\pi}^j}{(j-1)!}
                   = \frac{2^{n}{\pi}^{\frac{n(n+1)}{2}}}
                       {0! 1! \cdots (n-1)!}.
\end{equation}
Let us evaluate the volume of Grassmann manifold
$G_{k,n}(\fukuso)$:
\[
    G_{k,n}(\fukuso) \cong \frac{U(n)}{U(k)\times U(n-k)}\Longrightarrow
\mbox{Vol}(G_{k,n}(\fukuso)) = \frac{\mbox{Vol}(U(n))}
                              {\mbox{Vol}(U(k))\times \mbox{Vol}(U(n-k))}.
\]
%
         From (\ref{eq:unitary-volume}) we obtain
\begin{eqnarray}
   \label{eq:grassmann-volume}
  \mbox{Vol}(G_{k,n}(\fukuso)) &=&
       \frac{0!1! \cdots (k-1)!\ 0!1! \cdots (n-k-1)!}
            {0!1! \cdots \cdots (n-1)!}
       {\pi}^{k(n-k)} \nonumber \\
    &=& \frac{0!1! \cdots (k-1)!}{(n-k)! \cdots (n-2)!(n-1)!}
        {\pi}^{k(n-k)}.
\end{eqnarray}

\section{A Question}
\label{question}
\setcounter{equation}{0}
Combining (\ref{eq:grassmann-integral}) with (\ref{eq:grassmann-volume})
we have the main result
\begin{equation}
    \label{eq:identity}
   \int_{M(n-k,k;\fukuso)} \frac{\prod_{i,j} \frac{d{\bar z}_{ij}dz_{ij}}{2\ai}
         }{ \left\{\mbox{det}({\bf 1}_{k}+Z^{\dagger}Z)\right\}^{n} }\
   =\  \frac{0!1! \cdots (k-1)!}{(n-k)! \cdots (n-2)!(n-1)!}\,%
       {\pi}^{k(n-k)}.
\end{equation}
It has to be emphasized that the right hand side has been obtained by an 
indirect path.
  Is it really easy (or practical) to carry out the integration to obtain
the right hand side?  As far as we know, the integral has not been
calculated except for the case $k=1$.

Let us review the case $k=1$:
\begin{equation}
    \label{eq:special-identity}
   \int_{{\fukuso}^{n-1}} \frac{1}{\ (1+\sum_{j=1}^{n-1}|z_{j}|^{2})^{n}\ }
        \prod_{j=1}^{n-1} \frac{d{\bar z}_{j}dz_{j}}{2\ai}
   =\ \frac{{\pi}^{n-1}}{(n-1)!}\ .
\end{equation}
The proof is as follows. First let us make a change of variables:
\begin{equation}
  \label{eq:henkan}
         z_{j} = \sqrt{r_{j}}\mbox{e}^{\ai \theta_{j}}\quad \mbox{for}\
                  1 \leq j \leq n-1.
\end{equation}
Then we have easily
\[
    \frac{d{\bar z}_{j}dz_{j}}{2\ai} = \frac{1}{2}dr_{j}d\theta_{j}.
\]
\par
\noindent
Under this change of variables (\ref{eq:special-identity}) becomes
\[
   (\ref{eq:special-identity}) =
    \int_{0}^{2\pi}\!\!\int_{0}^{\infty}
    \frac{1}{(1+\sum_{j=1}^{n-1}r_{j})^{n}}
    \prod_{j=1}^{n-1} \frac{d\theta_{j}}{2}
    \prod_{j=1}^{n-1}dr_{j}
     = {\pi}^{n-1}\int_{0}^{\infty}
         \frac{1}{(1+\sum_{j=1}^{n-1}r_{j})^{n}}
         \prod_{j=1}^{n-1}dr_{j}.
\]
Here let us once more make a change of variables from $(r_{1}, \cdots,
r_{n-1})$ to $(\xi_{1}, \cdots, \xi_{n-1})$:
\begin{eqnarray}
  \label{eq:henkan-2}
     r_{1} &=& \xi_{1}(1 - \xi_{2}),  \nonumber \\
     r_{2} &=& \xi_{1}\xi_{2}(1 - \xi_{3}),  \nonumber \\
      \vdots  && \hspace{1.5cm}\vdots  \nonumber \\
     r_{n-2} &=& \xi_{1}\xi_{2}\cdots \xi_{n-2}(1 - \xi_{n-1}), \nonumber \\
     r_{n-1} &=& \xi_{1}\xi_{2}\cdots \xi_{n-2}\xi_{n-1}.  \nonumber
\end{eqnarray}
Conversely we have
\begin{eqnarray}
  \label{eq:henkan-3}
    \xi_{1} &=& r_{1}+ r_{2}+ \cdots + r_{n-2}+ r_{n-1},  \nonumber \\
    \xi_{2} &=& \frac{r_{2}+ \cdots + r_{n-2}+ r_{n-1}}
                     {r_{1}+ r_{2}+ \cdots + r_{n-2}+ r_{n-1}},  \nonumber \\
   \vdots     && \hspace{1.5cm}\vdots  \nonumber \\
    \xi_{n-2} &=& \frac{r_{n-2}+ r_{n-1}}{r_{n-3}+ r_{n-2}+ r_{n-1}},
              \nonumber \\
    \xi_{n-1} &=& \frac{r_{n-1}}{r_{n-2}+ r_{n-1}}, \nonumber \\[6pt]
  && \hspace{-12mm}0 \leq \xi_{1} < \infty,\quad
0 \leq \xi_{2}, \cdots, \xi_{n-1} \leq 1,\nonumber\\[6pt]
&&\hspace{-12mm}
\prod_{j=1}^{n-1}dr_{j}={\xi_{1}}^{n-2}{\xi_{2}}^{n-3}\cdots{\xi_{n-2}}
                             \prod_{j=1}^{n-1}d\xi_{j}.\nonumber
\end{eqnarray}
%
Under this change of variables (\ref{eq:special-identity})
becomes
\begin{eqnarray}
   (\ref{eq:special-identity}) &=& {\pi}^{n-1}
   \int_{0}^{\infty}\frac{ {\xi_{1}}^{n-2} }{ (1+\xi_{1})^{n} }d\xi_{1}
   \int_{0}^{1}{\xi_{2}}^{n-3}d\xi_{2} \cdots
   \int_{0}^{1}{\xi_{n-2}}d\xi_{n-2} \nonumber \\
  &=& {\pi}^{n-1}
   \int_{0}^{1}{\xi_{1}}^{n-2}d\xi_{1}
   \int_{0}^{1}{\xi_{2}}^{n-3}d\xi_{2} \cdots
   \int_{0}^{1}{\xi_{n-2}}d\xi_{n-2} \nonumber \\
  &=& {\pi}^{n-1} \frac{1}{n-1}\frac{1}{n-2}\cdots\frac{1}{2}
   = \frac{{\pi}^{n-1}}{(n-1)!} \quad \blacksquare
\end{eqnarray}
The direct proof of (\ref{eq:identity}) for $k=1$ is relatively easy as
shown above. But for $k\geq 2$ we do not know such a proof (a direct proof
may be very complicated). Therefore let us present
\begin{flushleft}
{\bf Problem}\quad  Give a direct proof to
\end{flushleft}
\[
  \int_{M(n-k,k;\fukuso)} \frac{\prod_{i,j} \frac{d{\bar z}_{ij}dz_{ij}}{2\ai}
         }{ \left\{\mbox{det}({\bf 1}_{k}+Z^{\dagger}Z)\right\}^{n} }\
  =\ \frac{0!1! \cdots (k-1)!}
          {(n-k)! \cdots (n-2)!(n-1)!} {\pi}^{k(n-k)}.
\]
\par
\vspace{3mm}

As for another approach to the above problem we refer to \cite{FKS}.
In this paper coherent states based on Grassmann manifolds have been
constructed.

\section{Quantum Computing}
\label{qcomp}
\setcounter{equation}{0}
Let us move to the main subject of Quantum Computing. The typical examples 
of quantum
algorithms up to now are
\begin{itemize}
    \item  Factoring algorithm of integers by P. Shor \cite{PS},
    \item  Quantum data-base searching algorithm by L. Grover \cite{LG}.
\end{itemize}
See \cite{AS} and \cite{RP}, or \cite{LPS}  for general introduction.
\cite{RJ} and \cite{RJ2} are also recommended.

In Quantum Computing
we in general expect an exponential speedup compared to classical ones, so
we must construct necessarily unitary matrices in $U(n)$ in an efficient
manner when $n$ is a huge number like $2^{100}$.
\begin{flushleft}
{\bf Problem}\quad  How can we construct unitary matrices
  in an efficient manner?
\end{flushleft}
\par \noindent
Let us come back to (\ref{eq:grassmann-definition}). We denote the set
of $n\times n$ projection operators by
\begin{equation}
   \label{eq:total-grassmann}
   G_{n}(\fukuso)=\{P \in M(n;\fukuso) |\ P^2=P,\ P^{\dagger}=P \}.
\end{equation}
The elements of $G_{n}(\fukuso)$ are classified by the trace, so
$G_{n}(\fukuso)$ can be decomposed into a disjoint union
\begin{equation}
   \label{eq:union}
   G_{n}(\fukuso)=\ \bigcup_{k=0}^{n} G_{k,n}(\fukuso).
\end{equation}
For a $k$-dimensional subspace $V$ in $\fukuso^n$ ($0\leq k \leq n$) let $
\{ {\bf v}_{1},{\bf v}_{2}, \cdots , {\bf v}_{k} \}$ be an orthonormal
basis (namely, $<{\bf v}_{i},{\bf v}_{j}> = \delta_{ij}$) and set
\begin{equation}
   \label{eq:stiefel}
   V = ({\bf v}_{1},{\bf v}_{2}, \cdots , {\bf v}_{k}) \in M(n,k;\fukuso).
\end{equation}
We have identified a $k$-dimensional subspace $V$ with a matrix $V$ in
(\ref{eq:stiefel}) for simplicity (maybe there is no confusion).
Then we have an equivalence
\[
  \{ {\bf v}_{1},{\bf v}_{2}, \cdots , {\bf v}_{k} \} :
  \mbox{orthonormal}\  \Leftrightarrow\  V^{\dagger}V = {\bf 1}_{k}.
\]
Then it is easy to see that all orthonormal basis in $V$  are given by
\begin{equation}
    \{Va |\ a \in U(k) \}.
\end{equation}
The projection corresponding to $V$ is written by
\begin{equation}
    P = VV^{\dagger} \in G_{k,n}(\fukuso).
\end{equation}
We remark that $(Va)(Va)^{\dagger}=Va{a}^{\dagger}V^{\dagger}=
VV^{\dagger}=P$, namely $P$ is of course independent of $a \in U(k)$.
This $P$ is also expressed as
\begin{equation}
   \label{eq:summation-1}
    P = \sum_{j=1}^{k} {\bf v}_{j}{{\bf v}_{j}}^{\dagger}.
\end{equation}
If we use  Dirac's bra-ket notation ${\bf v}_{j}=\ket{j}$, then
$
    P = \sum_{j=1}^{k} \ket{j}\bra{j}
$.
This notation may be popular in Physics rather than (\ref{eq:summation-1}).

How can we construct an element of unitary group from an element of
Grassmann manifolds? We have a canonical method, namely
\begin{equation}
   \label{eq:uniton}
    G_{n}(\fukuso) \longrightarrow U(n) : P \longmapsto
    U = \mbox{1}_{n} - 2P.
\end{equation}
This $U$ is called a uniton in the field of harmonic maps.
Moreover we can consider a product of some unitons, namely,
for any $S \subset \{0,1, \cdots, n-1, n \}$
\begin{equation}
   \label{eq:uniton-many}
    U = \prod_{j\in S} (\mbox{1}_{n} - 2P_{j})\quad \mbox{for}\
    P_{j} \in G_{j,n}(\fukuso).
\end{equation}
In particular
\begin{equation}
   \label{eq:uniton-full}
    U = \prod_{j=1}^{n-1} (\mbox{1}_{n} - 2P_{j})\quad \mbox{for}\
    P_{j} \in G_{j,n}(\fukuso)
\end{equation}
is very important in the field of harmonic maps, see for example
\cite{GD} and \cite{WJ}.

\par \noindent
Many important unitary matrices are made by this way%
\footnote{Those used in \cite{GZ} for data-base searching
algorithms are of  this form with appropriate $P_{j}$.}.

\vspace{3mm}
\par \noindent
These unitary matrices also play an important role in Quantum
Computing as shown in the following.

We consider a qubit ({\bf qu}antum {\bf bit}) space of quantum particles.
The 1-qubit space is identified with $\fukuso^{2}$ with basis
$\{\ket{0}, \ket{1} \}$ ;
\[
    \fukuso^{2} = \mbox{Vect}_{\fukuso}\{\ket{0}, \ket{1} \},\quad
    \ket{0}= {1\choose 0} ,\quad \ket{1}= {0\choose 1}\ .
\]
The qubit space of $t$-particles is the {\bf tensor product} (not direct
sum!) of $\fukuso^{2}$
\begin{equation}
   \label{eq:tensor-space}
   \fukuso^{2}\otimes \fukuso^{2} \otimes \cdots \otimes \fukuso^{2}
   \equiv (\fukuso^{2})^{\otimes t}
\end{equation}
with basis
\[ \left\{
    \ket{{n_{1},n_{2}, \dots, n_{t}}} = \ket{n_{1}}\otimes \ket{n_{2}}\otimes
     \cdots \otimes \ket{n_{t}}\
     |\  n_{j} \in \integer_{2} = \{0,1 \}\
    \right\}.
\]
For example,
\[
    \ket{0}\otimes \ket{0}=
   \left(
     \begin{array}{c}
       1 \\
       0 \\
       0 \\
       0
     \end{array}
   \right),\
    \ket{0}\otimes \ket{1}=
   \left(
     \begin{array}{c}
       0 \\
       1 \\
       0 \\
       0
     \end{array}
   \right),\
    \ket{1}\otimes \ket{0}=
   \left(
     \begin{array}{c}
       0 \\
       0 \\
       1 \\
       0
     \end{array}
   \right),\
    \ket{1}\otimes \ket{1}=
   \left(
     \begin{array}{c}
       0 \\
       0 \\
       0 \\
       1
     \end{array}
   \right).
\]
\vspace{5mm}

Now we take the Walsh-Hadamard transformation $W$ defined by
\begin{equation}
   W : \ket{0} \longrightarrow \frac{1}{\sqrt{2}}(\ket{0}+ \ket{1}), \quad
   W : \ket{1} \longrightarrow \frac{1}{\sqrt{2}}(\ket{0}- \ket{1}),
\end{equation}
in matrix notation,
\begin{equation}
   \label{eq:w-a}
   W = \frac{1}{\sqrt{2}}
     \left(
        \begin{array}{rr}
            1& 1 \\
            1& -1
        \end{array}
     \right)\ \in \ O(2)\ \subset U(2).
\end{equation}
This transformation (or matrix) is  unitary  and it
plays a very important role in Quantum Computing. Moreover is easy to
realize it in Quantum Optics.
Let us list some important properties of $W$:
\begin{eqnarray}
      &&W^{2}={\bf 1}_{2},\ \ W^{\dagger}=W=W^{-1}, \\
      &&\sigma_{1}= W\sigma_{3}W^{-1},
\end{eqnarray}
where $\{\sigma_{1}, \sigma_{2}, \sigma_{3} \}$ are the Pauli matrices:
\begin{equation}
\label{eq:pauli-matrices}
   \sigma_{1}=
     \left(
        \begin{array}{cc}
             & 1 \\
            1&
        \end{array}
     \right),\
   \sigma_{2}=
     \left(
        \begin{array}{cc}
               & -\ai \\
            \ai&
        \end{array}
     \right),\
   \sigma_{3}=
     \left(
        \begin{array}{cc}
            1&  \\
             &-1
        \end{array}
     \right) .
\end{equation}
See Appendix {\bf B} for a generalization of Pauli matrices.
Next we consider $t$-tensor product of $W$ ($t$ $\in \futon$) :
\begin{equation}
    \label{eq:t-tensor}
       W^{{\otimes} t}=W\otimes W\otimes \cdots \otimes W \ (
       t \mbox{-times}) .
\end{equation}
This matrix of course operates on the space (\ref{eq:tensor-space}).
For example
\begin{equation}
   \label{eq:2-tensor}
  W\otimes W = \frac{1}{2}
   \left(
     \begin{array}{rrrr}
          1&  1&  1&  1  \\
          1& -1&  1& -1  \\
          1&  1& -1& -1  \\
          1& -1& -1&  1
     \end{array}
   \right),
\end{equation}
and
\begin{equation}
   \label{eq:3-tensor}
  W\otimes W\otimes W  = \frac{1}{\sqrt{8}}
   \left(
     \begin{array}{rrrrrrrr}
          1&  1&  1&  1&  1&  1&  1&  1 \\
          1& -1&  1& -1&  1& -1&  1& -1 \\
          1&  1& -1& -1&  1&  1& -1& -1 \\
          1& -1& -1&  1&  1& -1& -1&  1 \\
          1&  1&  1&  1& -1& -1& -1& -1 \\
          1& -1&  1& -1& -1&  1& -1&  1 \\
          1&  1& -1& -1& -1& -1&  1&  1 \\
          1& -1& -1&  1& -1&  1&  1& -1
     \end{array}
   \right).
\end{equation}
\vspace{3mm}

Hereafter we  set $n = 2^{t}$. Then (\ref{eq:t-tensor}) means
$W^{{\otimes} t} \in U(n)$. The very important fact is that
(\ref{eq:t-tensor}) can be constructed by only $t(= \mbox{log}_{2}(n))$-steps
in Quantum Computing. Let us show the matrix-component of 
(\ref{eq:t-tensor}) is given by
\begin{equation}
   \bra{{i_{1},i_{2}, \dots, i_{t}}}W^{{\otimes} t}
   \ket{{j_{1},j_{2}, \dots, j_{t}}} = \frac{1}{\sqrt{n}}
   (-1)^{\sum_{k=1}^{t} i_{k}j_{k}}
\end{equation}
or if we set
\[
  |i) = \ket{i_{1}}\otimes \ket{i_{2}}\otimes \cdots \otimes \ket{i_{t}},
  \quad
  i = i_{1}2^{t-1}+i_{2}2^{t-2}+ \cdots + i_{t},\quad 0 \leq i \leq n-1
\]
we have
\begin{equation}
   \label{eq:inner-product}
   (i|W^{{\otimes} t}|j) = \frac{1}{\sqrt{n}}(-1)^{i\cdot j}
\end{equation}
where $i\cdot j$ means the sum of bit-wise products $\sum_{k=1}^{t} 
i_{k}j_{k}$.
\par
\noindent
The proof goes as follows. From (\ref{eq:w-a}) we know
\[
   W\ket{i}=\frac{1}{\sqrt{2}} (\ket{0}+(-1)^{i}\ket{1})
           =\frac{1}{\sqrt{2}} \sum_{k=0}^{1}(-1)^{ik}\ket{k},
\]
which implies
\begin{eqnarray}
   \label{eq:operation}
     W^{{\otimes} t}|j) &=& W\ket{j_{1}}\otimes W\ket{j_{2}}\otimes \cdots
                            \otimes W\ket{j_{t}}  \nonumber \\
      &=& \frac{1}{\sqrt{2^{t}}}\sum_{k_{1}=0}^{1}\sum_{k_{2}=0}^{1}\cdots
          \sum_{k_{t}=0}^{1}(-1)^{k_{1}j_{1}+k_{2}j_{2}+\cdots +k_{t}j_{t}}
          \ket{k_{1}}\otimes \ket{k_{2}}\otimes \cdots \ket{k_{t}}
          \nonumber \\
      &=& \frac{1}{\sqrt{n}}\sum_{k=0}^{n-1}(-1)^{k\cdot j}|k).
\end{eqnarray}
Therefore we obtain
\[
     (i|W^{{\otimes} t}|j)
       = \frac{1}{\sqrt{n}}\sum_{k=0}^{n-1}(-1)^{k\cdot j}(i|k)
       = \frac{1}{\sqrt{n}}\sum_{k=0}^{n-1}(-1)^{k\cdot j}{\delta}_{ik}
       = \frac{1}{\sqrt{n}}(-1)^{i\cdot j} \quad \blacksquare
\]

\par \noindent
Moreover (\ref{eq:t-tensor}) has an interesting property which
we can guess from (\ref{eq:2-tensor}) and (\ref{eq:3-tensor}):
\begin{equation}
    \sum_{j=0}^{n-1} (i|W^{{\otimes} t}|j) =
    \left\{
        \begin{array}{rl}
            \sqrt{n},& \quad \mbox{if}\ \  i = 0 \\
             0,& \quad \mbox{othewise}
        \end{array}
    \right.
\end{equation}
or
\begin{equation}
    \sum_{i=0}^{n-1} (i|W^{{\otimes} t}|j) =
    \left\{
        \begin{array}{rl}
            \sqrt{n},& \quad \mbox{if}\ \  j = 0 \\
             0,& \quad \mbox{othewise}.
        \end{array}
    \right.
\end{equation}

Let us clarify the meaning of (\ref{eq:inner-product}) from the point of
view of Group Theory.

\par \noindent
We note that $\integer_{2}$ is an abelian group with operation  $\oplus$
\begin{equation}
   \label{eq:oplus}
   0\oplus 0=0,\ 0\oplus 1=1,\ 1\oplus 0=1,\ 1\oplus 1=0.
\end{equation}
Then ${{\integer}_{2}}^{t}$ is a natural product group of $\integer_{2}$.
We denote its element by
\[
   {\bf i}=(i_{1},i_{1}, \cdots, i_{t}) \longleftrightarrow
   i = i_{1}2^{t-1}+i_{2}2^{t-2}+ \cdots + i_{t}.
\]
\par \noindent
For ${\bf i}\in {{\integer}_{2}}^{t}$ we define
\begin{equation}
    {\chi}_{{\bf i}} : {{\integer}_{2}}^{t} \longrightarrow \fukuso^{*}=
    \fukuso-\{0 \}, \quad
    {\chi}_{{\bf i}}({\bf j})=\sqrt{n}(i|W^{{\otimes} t}|j)=(-1)^{i\cdot j}.
\end{equation}
Then we can show that
\begin{equation}
   \label{eq:character}
    {\chi}_{{\bf i}}({\bf j}\oplus {\bf k})=
           {\chi}_{{\bf i}}({\bf j}){\chi}_{{\bf i}}({\bf k}).
\end{equation}
That is, ${\chi}_{{\bf i}}$ is a character of the abelian group
  ${{\integer}_{2}}^{t}$.
\par \noindent
The proof is as follows.\quad From (\ref{eq:oplus}) we know
\begin{equation}
\label{eq:mod2-identity}
   x\oplus y = x + y -2xy \quad \mbox{for}\ \ x, y \in \integer_{2}.
\end{equation}
Therefore we obtain
\begin{eqnarray}
  {\chi}_{{\bf i}}({\bf j}\oplus {\bf k})
    &=&  (-1)^{\sum_{l=1}^{t} i_{l}(j_{l}\oplus k_{l})}
     =  (-1)^{\sum_{l=1}^{t} i_{l}(j_{l}+k_{l}-2j_{l}k_{l})}
     =  (-1)^{\sum_{l=1}^{t} i_{l}(j_{l}+k_{l})} \nonumber \\
    &=&  (-1)^{\sum_{l=1}^{t} i_{l}j_{l}}
         (-1)^{\sum_{l=1}^{t} i_{l}k_{l}}
     =  (-1)^{i\cdot j}(-1)^{i\cdot k}
     = {\chi}_{{\bf i}}({\bf j}){\chi}_{{\bf i}}({\bf k}) \quad \blacksquare
\end{eqnarray}
These characters play an important role in Discrete Fourier Transform,
see \cite{RJ} or recent preprint \cite{RJ2}.

Now we consider a controlled NOT operation (gate) which we will denote by
C-NOT in the following. It is defined by
\begin{eqnarray}
\mbox{C-NOT} : &\quad&
   \kett{0}{0}\to \kett{0}{0},\quad \kett{0}{1}\to \kett{0}{1}, \nonumber \\
&&
  \kett{1}{0}\to \kett{1}{1},\quad  \kett{1}{1}\to \kett{1}{0}
\end{eqnarray}
or more compactly
\begin{equation}
\mbox{C-NOT} : \ket{a,b}\longrightarrow \ket{a,a\oplus b}, \quad
a,\ b \in \integer_{2}.
\end{equation}
Graphically it is expressed as
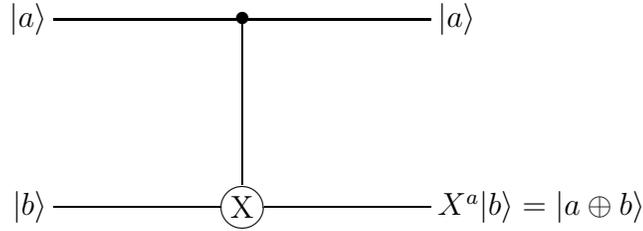
\begin{center}
\setlength{\unitlength}{1mm}
\begin{picture}(160,35)
\put(50,30){\line(1,0){50}}   
\put(50,5){\line(1,0){22}}   
\put(78,5){\line(1,0){22}}   
\put(40,25){\makebox(9,10)[r]{$|a\rangle$}} 
\put(40,0){\makebox(9,10)[r]{$|b\rangle$}} 
\put(101,25){\makebox(9,10)[l]{$|a\rangle$}} 
\put(101,0){\makebox(9,10)[l]{$X^{a}\ket{b}=|a\oplus b\rangle$}}
\put(75,8){\line(0,1){22}}     
\put(72,25){\makebox(6,10){$\bullet$}} 
\put(75,5){\circle{6}}               
\put(72,0){\makebox(6,10){X}}         
\end{picture}
\end{center}
and the matrix representation is
\begin{eqnarray}
   \label{eq:c-not}
\mbox{C-NOT}=
\left(
   \begin{array}{cccc}
        1&0&0&0 \\
        0&1&0&0 \\
        0&0&0&1 \\
        0&0&1&0
   \end{array}
\right).
\end{eqnarray}
Here $a\oplus b=a+b\ (\mbox{mod}\ 2)$ and we note the relation
\begin{equation}
X^{a}\ket{b}\equiv \sigma_{1}^{a}\ket{b}=\ket{a\oplus b}
\quad \mbox{for}\quad a,\ b \in \integer_{2}.
\end{equation}

\par \noindent
In this case the first bit is called a control bit and the
second a target bit.

\par \noindent
Of course we can consider the reverse case. Namely,
the first bit is a target one and the second a control one, which is also
called the controlled NOT operation:
\begin{equation}
\mbox{C-NOT} : \ket{a,b}\longrightarrow \ket{a\oplus b,b},
\quad a,\ b \in \integer_{2},
\end{equation}
and the matrix representation is
\begin{eqnarray}
   \label{eq:c-hat-not}
\mbox{C-NOT}=
\left(
   \begin{array}{cccc}
        1&0&0&0 \\
        0&0&0&1 \\
        0&0&1&0 \\
        0&1&0&0
   \end{array}
\right).
\end{eqnarray}

\par \noindent
A comment is in order. In the 1-qubit case we may assume that we can
construct all unitary operations in $U(2)$ (we call the operation
universal). In the 2-qubit case how can we construct all unitary
operations in $U(4)$?  If we can construct the C-NOT (\ref{eq:c-not}),
(\ref{eq:c-hat-not}) in our system,
then we can show the operation is universal, see \cite{3th} and \cite{9th}.
This is a crucial point in quantum computing.
We comment here that the C-NOT (\ref{eq:c-not}) can be written as a uniton 
(\ref{eq:uniton})
\begin{equation}
    \mbox{C-NOT} = {\bf 1}_{4}- 2P
\end{equation}
where
\begin{equation}
    P = \frac{1}{2}
    \left(
   \begin{array}{cccc}
        0& 0&  0&  0 \\
        0& 0&  0&  0 \\
        0& 0&  1& -1 \\
        0& 0& -1&  1
   \end{array}
\right),
\end{equation}
and this $P$ can be diagonalized by making use of Walsh-Hadamard
transformation (\ref{eq:w-a}) like
\begin{equation}
    P = ({\bf 1}_{2}\otimes W){\widetilde E_{1}}
        ({\bf 1}_{2}\otimes W)^{-1}
      = ({\bf 1}_{2}\otimes W)(\sigma_{1}\otimes \sigma_{1})E_{1}
        (\sigma_{1}\otimes \sigma_{1})^{-1}({\bf 1}_{2}\otimes W)^{-1}
\end{equation}
where
\begin{equation}
  {\widetilde E_{1}}=
    \left(
   \begin{array}{cccc}
        0&  &   &   \\
         & 0&   &   \\
         &  &  0&   \\
         &  &   &  1
   \end{array}
\right), \qquad
  E_{1}=
    \left(
   \begin{array}{cccc}
        1&  &   &   \\
         & 0&   &   \\
         &  &  0&   \\
         &  &   &  0
   \end{array}
\right).
\end{equation}
%

More generally for the $t$-qubit case
we can construct ($t-1$)-repeated controlled-not operator and show it is
a uniton.

The $(t-1)$-repeated controlled-not operation is defined by
\begin{eqnarray}
\label{eq:repeated-controlled-not}
\mbox{C}^{(t-1)}\mbox{-NOT} : \ket{a_1,a_2,\cdots,a_{t-1},a_t}
\longrightarrow &&
\ket{a_1,a_2,\cdots,a_{t-1},a_{1}a_{2}\cdots a_{t-1}\oplus a_t}, \nonumber \\
&& a_{k}\in \integer_{2}\quad (k=1,2,\cdots, t),
\end{eqnarray}
or in matrix form
\begin{equation}
  \mbox{C}^{(t-1)}\mbox{-NOT}=\left(
       \begin{array}{cccccc}
         1&     &      &  &  &   \\
          &\cdot&      &  &  &   \\
          &     &\cdot &  &  &   \\
          &     &      &1 &  &   \\
          &     &      &  &0 &1   \\
          &     &      &  &1 &0   \\
       \end{array}
      \right):\quad 2^{t}\times 2^{t}-\mbox{matrix}.
\end{equation}

As for the explicit construction of ($t-1$)-repeated controlled-not
operator see \cite{9th} or \cite{KF6}. See also Appendix.
But unfortunately the construction is not {\bf efficient} !

\par \noindent A comment is in order.
In \cite{9th} a rough estimation of the number of steps to construct
the operator (\ref{eq:repeated-controlled-not}) is given and it is
confirmed that an efficient construction is possible. But no explicit
construction is  given.

\par \vspace{3mm} \noindent
By the way,  since
\begin{equation}
   {{\bf 1}_{2}}^{\otimes (t-1)}\otimes W
      =\left(
       \begin{array}{ccccc}
         W&     &      &  &     \\
          &\cdot&      &  &     \\
          &     &\cdot &  &     \\
          &     &      &W &     \\
          &     &      &  &W    \\
       \end{array}
      \right),
\end{equation}
we have
\begin{eqnarray}
\label{eq:transformation}
   \left({{\bf 1}_{2}}^{\otimes (t-1)}\otimes W \right)
   \mbox{C}^{(t-1)}\mbox{-NOT}
   \left({{\bf 1}_{2}}^{\otimes (t-1)}\otimes W \right)
      &=&\left(
       \begin{array}{cccccc}
         1&     &      &  &  &   \\
          &\cdot&      &  &  &   \\
          &     &\cdot &  &  &  \\
          &     &      &1 &  &   \\
          &     &      &  &1 &   \\
          &     &      &  &  &-1  \\
       \end{array}
      \right)  \nonumber \\
      &=& {\bf 1}_{n}-2|n-1)(n-1|.
\end{eqnarray}
Therefore the construction of $\mbox{C}^{(t-1)}\mbox{-NOT}$ and
${\bf 1}_{n}-2|n-1)(n-1|$ have almost the same  number of steps.
Is it possible to construct this operator efficiently ?

\par \noindent
As far as we know an explicit and efficient construction of this operation
has not yet been given\footnote{I have not yet succeeded in such a 
construction.}.
\begin{flushleft}
{\bf Problem}\quad  Give an explicit and efficient algorithm to this
operation.
\end{flushleft}

Let us consider a set
\begin{equation}
       \{ F_{k}\ |\ 1 \leq k \leq n \}
\end{equation}
where
\[
   F_{k}={\bf 1}_{n}-2E_{k}=
    \left(
   \begin{array}{cc}
        -{\bf 1}_{k}&     \\
         & {\bf 1}_{n-k}
   \end{array}
\right).
\]
If $F_{1}$ can be constructed, then the other $F$'s can be easily obtained.
First let us show this with a simple example ($t$=2) :
\[
F_{1}=
    \left(
   \begin{array}{cccc}
       -1&  &   &   \\
         & 1&   &   \\
         &  &  1&   \\
         &  &   &  1
   \end{array}
\right)
={\bf 1}_{4}-2|0)(0|.
\]
Now we set
\begin{eqnarray}
   U_{1} &=& {\bf 1}_{2}\otimes \sigma_{1}=
    \left(
   \begin{array}{cccc}
         & 1&   &   \\
        1&  &   &   \\
         &  &   & 1 \\
         &  &  1&
   \end{array}
\right), \quad
   U_{2} = \sigma_{1}\otimes {\bf 1}_{2}=
    \left(
   \begin{array}{cccc}
         &  &  1&   \\
         &  &   & 1 \\
        1&  &   &   \\
         & 1&   &
   \end{array}
\right), \nonumber \\
   U_{3} &=& \sigma_{1}\otimes \sigma_{1}=
    \left(
   \begin{array}{cccc}
         &  &   &1  \\
         &  &  1&   \\
         & 1&   &   \\
        1&  &   &
   \end{array}
\right), \nonumber
\end{eqnarray}
then we have
\begin{eqnarray}
   U_{1}F_{1}U_{1}&=&
    \left(
   \begin{array}{cccc}
        1&  &   &   \\
         &-1&   &   \\
         &  &  1&   \\
         &  &   &  1
   \end{array}
\right)
={\bf 1}_{4}-2|1)(1|, \nonumber \\
   U_{2}F_{1}U_{2}&=&
    \left(
   \begin{array}{cccc}
        1&  &   &   \\
         & 1&   &   \\
         &  & -1&   \\
         &  &   &  1
   \end{array}
\right)
={\bf 1}_{4}-2|2)(2|, \nonumber \\
   U_{3}F_{1}U_{3}&=&
    \left(
   \begin{array}{cccc}
        1&  &   &   \\
         & 1&   &   \\
         &  &  1&   \\
         &  &   &  -1
   \end{array}
\right)
={\bf 1}_{4}-2|3)(3|, \nonumber
\end{eqnarray}
so that it is easy to check
\begin{equation}
       F_{1}\left(U_{1}F_{1}U_{1} \right)=F_{2},\
       F_{2}\left(U_{2}F_{1}U_{2} \right)=F_{3}\ \ \mbox{and}\ \
       F_{3}\left(U_{3}F_{1}U_{3} \right)=-{\bf 1}_{4}.
\end{equation}
Let us prove the general case.
For $i = i_{1}2^{t-1}+i_{2}2^{t-2}+ \cdots + i_{t}$ $(0\leq i \leq n-1)$
we set
\begin{equation}
    U_{i}=\sigma_{1}^{i_{1}}\otimes \sigma_{1}^{i_{2}}\otimes \cdots
          \otimes \sigma_{1}^{i_{t}},
          \quad (U_{i}^{\dagger}=U_{i}=U_{i}^{-1}).
\end{equation}
Since
\begin{eqnarray}
    U_{i}|0)&=&
    \sigma_{1}^{i_{1}}\otimes \sigma_{1}^{i_{2}}\otimes \cdots
          \otimes \sigma_{1}^{i_{t}}\ (\ket{0}\otimes \ket{0}\otimes
          \cdots \otimes \ket{0})  \nonumber \\
    &=& \sigma_{1}^{i_{1}}\ket{0}\otimes \sigma_{1}^{i_{2}}\ket{0}\otimes
         \cdots
         \otimes \sigma_{1}^{i_{t}}\ket{0}  \nonumber \\
    &=& \ket{i_{1}}\otimes \ket{i_{2}}\otimes \cdots \otimes \ket{i_{t}}
    \equiv |i),
\end{eqnarray}
we have
\begin{equation}
   \label{eq:ufu}
   {\bf 1}_{n}- 2|i)(i| = U_{i}\left(\vTm {\bf 1}_{n}- 2|0)(0| \right)U_{i}
                        = U_{i}F_{1}U_{i}.
\end{equation}
Therefore it is easy to see that
\begin{equation}
       F_{k}\left(U_{k}F_{1}U_{k} \right)=F_{k+1}\quad
       (1\leq k \leq n-1). \quad \blacksquare
\end{equation}
We note that this procedure is not efficient.

\vspace{3mm}
Now, let us make a comment on Grover's data-base searching algorithm.
In his algorithm the following two unitary operations
play an essential role:
\begin{equation}
\label{eq:two-unitary-operations}
    {\bf 1}_{n}- 2|i)(i|,\quad  {\bf 1}_{n}- 2|s)(s|,
\end{equation}
in which the state $|s)$ ($s$ stands for {\em sum\/}) is defined by
\begin{equation}
   |s)\equiv \frac{1}{\sqrt{n}} \sum_{i=0}^{n-1} |i) = W^{\otimes t}|0).
\end{equation}
We find via (\ref{eq:operation}) that
\begin{equation}
   {\bf 1}_{n}- 2|s)(s|= W^{\otimes t}\left(\vTm {\bf 1}_{n}- 2|0)(0|\right)
                      W^{\otimes t}= W^{\otimes t}F_{1}W^{\otimes t}.
\end{equation}
Namely, the two operations (\ref{eq:two-unitary-operations}) are both
unitons and can be diagonalized by the efficient unitary operations
$U_{i}$ and $ W^{\otimes t}$.

\vspace{3mm}
Finally, let us mention about a relation between $(t-1)$-repeated controlled-
not operation and $F_{1}$. Since
\[
\left({{\bf 1}_{2}}^{\otimes (t-1)}\otimes W \right)
\mbox{C}^{(t-1)}\mbox{-NOT}
\left({{\bf 1}_{2}}^{\otimes (t-1)}\otimes W \right)
=
{\bf 1}_{n}-2|n-1)(n-1|
=
U_{n-1}F_{1}U_{n-1}
\]
by (\ref{eq:transformation}), we have
\[
    F_{1}= U_{n-1}\left({{\bf 1}_{2}}^{\otimes (t-1)}\otimes W \right)
           \mbox{C}^{(t-1)}\mbox{-NOT}
           \left({{\bf 1}_{2}}^{\otimes (t-1)}\otimes W \right) U_{n-1}.
\]
By substituting
$U_{n-1}=\sigma_{1}\otimes \sigma_{1}\otimes \cdots \otimes
\sigma_{1}\otimes \sigma_{1}$
into the equation above we arrive at the desired relation
\begin{equation}
  F_{1}=\left({\sigma_{1}}^{\otimes (t-1)}\otimes \sigma_{1}W \right)
        \mbox{C}^{(t-1)}\mbox{-NOT}
        \left({\sigma_{1}}^{\otimes (t-1)}\otimes W\sigma_{1} \right).
\end{equation}
%

\section{Holonomic Quantum Computation}
\label{holonomy}
\setcounter{equation}{0}

In this section we briefly introduce a simplified version
  of Holonomic Quantum Computation.
The full  story would require detailed knowledge of Quantum Mechanics,
Quantum Optics and Global Analysis, so it will
have to wait for another occasion.

This model was proposed by Zanardi and Rasetti \cite{ZR} and \cite{PZR}
and it has been developed by Fujii \cite{KF1}, \cite{KF2}, \cite{KF3},
\cite{KF4} and Pachos \cite{PC}, \cite{PZ}.

This model uses the non-abelian Berry phase (quantum holonomy in the
mathematical terminology \cite{MN}) in the process of quantum computing.
In this model a Hamiltonian (including some parameters) must have
certain degeneracy because an adiabatic connection (the non-abelian Berry
connection) is introduced in terms of the degeneracy, see \cite{SW}.
In other words, a quantum computational bundle is introduced on some
parameter space due to this degeneracy and the canonical connection of
this bundle is just the one above.

\par \noindent
On this bundle Holonomic Quantum Computation is performed by making use of
the holonomy operations. We note our method is completely geometrical.

\vspace{5mm}
Here we introduce {\bf quantum computational bundles}, \cite{KF1}, \cite{KF2},
and \cite{KF4}.
For this purpose we need universal principal and vector bundles over
infinite dimensional Grassmann manifolds. We also need an
infinite dimensional vector space called a Hilbert (or Fock) space.

Let $\calh$ be a separable Hilbert space over $\fukuso$.
For $m\in{\bf N}$, we set
\begin{equation}
   \label{eq:stmh}
   \kansu{\stm}{\cal H}
   \equiv
   \left\{
     V=\left( v_1,\cdots,v_m\right)
     \in
     \calh\times\cdots\times\calh\vert\ V^\dagger V=1_m\right\},
\end{equation}
where $1_m$ is a unit matrix in $\kansu{M}{m,\fukuso}$.
This is called a (universal) Stiefel manifold.
Note that the unitary group $U(m)$ acts on $\kansu{\stm}{\calh}$
  from the right:
\begin{equation}
   \label{eq:stmsha}
   \kansu{\stm}{\calh}\times\kansu{U}{m}
   \rightarrow
   \kansu{\stm}{\calh}: \left( V,a\right)\mapsto Va.
\end{equation}
Next we define a (universal) Grassmann manifold
\begin{equation}
   \kansu{\grm}{\calh}
   \equiv
   \left\{
     X\in\kansu{M}{\calh}\vert\
     X^2=X, X^\dagger=X\  \mathrm{and}\  \mathrm{tr}X=m\right\},
\end{equation}
where $M(\calh)$ denotes a space of all bounded linear operators on $\calh$.
Then we have a projection
\begin{equation}
   \label{eq:piteigi}
   \pi : \kansu{\stm}{\calh}\rightarrow\kansu{\grm}{\calh},
   \quad \kansu{\pi}{V}\equiv
   VV^{\dagger}=\sum_{j=1}^{m}v_{j}v_{j}^{\dagger},
\end{equation}
compatible with the action (\ref{eq:stmsha})
($\kansu{\pi}{Va}=Va(Va)^\dagger=Vaa^\dagger V^\dagger=VV^\dagger=
\kansu{\pi}{V}$).

Now the set
\begin{equation}
   \label{eq:principal}
   \left\{
     \kansu{U}{m}, \kansu{\stm}{\calh}, \pi, \kansu{\grm}{\calh}
   \right\},
\end{equation}
is called a (universal) principal $U(m)$ bundle,
see \cite{MN} and \cite{KF1}.\quad We set
\begin{equation}
   \label{eq:emh}
   \kansu{\eem}{\cal H}
   \equiv
   \left\{
     \left(X,v\right)
     \in
     \kansu{\grm}{\calh}\times\calh \vert\ Xv=v \right\}.
\end{equation}
Then we have also a projection
\begin{equation}
   \label{eq:piemgrm}
   \pi : \kansu{\eem}{\calh}\rightarrow\kansu{\grm}{\calh}\ ,
   \quad \kansu{\pi}{\left(X,v\right)}\equiv X.
\end{equation}
The set
\begin{equation}
   \label{eq:universal}
   \left\{
     \fukuso^m, \kansu{\eem}{\calh}, \pi, \kansu{\grm}{\calh}
   \right\},
\end{equation}
is called a (universal) $m$-th vector bundle. This vector bundle is
associated with the principal $U(m)$ bundle (\ref{eq:principal}).

Next let $M$ be a finite or infinite dimensional differentiable manifold
and the map $P : M \rightarrow \kansu{\grm}{\calh}$ be given (called a
projector). Using this $P$ we can define the  pullback bundles over $M$
from (\ref{eq:principal}) and (\ref{eq:universal}):
\begin{eqnarray}
   \label{eq:hikimodoshi1}
   &&\left\{\kansu{U}{m},\widetilde{St}, \pi_{\widetilde{St}}, M\right\}
   \equiv
   P^*\left\{\kansu{U}{m}, \kansu{\stm}{\calh}, \pi,
   \kansu{\grm}{\calh}\right\}, \\
   \label{eq:hikimodoshi2}
   &&\left\{\fukuso^m,\widetilde{E}, \pi_{\widetilde{E}}, M\right\}
   \equiv
   P^*\left\{\fukuso^m, \kansu{\eem}{\calh}, \pi, \kansu{\grm}{\calh}\right\},
\end{eqnarray}
see \cite{MN}. Of course the second bundle (\ref{eq:hikimodoshi2}) is  a 
vector bundle
associated with the first one (\ref{eq:hikimodoshi1}):
\vspace{3mm}
\[
    \matrix{
     \kansu{U}{m}&&\kansu{U}{m}\cr
     \Big\downarrow&&\Big\downarrow\cr
     \widetilde{St}&\longrightarrow&\kansu{\stm}{\calh}\cr
     \Big\downarrow&&\Big\downarrow\cr
     {M}&\stackrel{P}{\longrightarrow}&\kansu{\grm}{\calh}\cr
            } \qquad \qquad
    \matrix{
     \fukuso^m&&\fukuso^m\cr
     \Big\downarrow&&\Big\downarrow\cr
     \widetilde{E}&\longrightarrow&\kansu{E_m}{\calh}\cr
     \Big\downarrow&&\Big\downarrow\cr
     {M}&\stackrel{P}{\longrightarrow}&\kansu{\grm}{\calh}\cr
            }
\]
\par \vspace{3mm}
Let $\calm$ be a parameter space (a complex manifold in general)
and we denote by $\lam$ its element.
Let $\slam$ be a fixed reference point of $\calm$. Let $H_\lam$ be
a family of Hamiltonians parameterized by $\calm$  acting on the Fock space
$\calh$. We set $H_0$ = $H_\slam$ for simplicity and assume that this has
an $m$-fold degenerate vacuum:
\begin{equation}
   H_{0}v_j = \mathbf{0},\quad j = 1,\ldots, m.
\end{equation}
These $v_j$'s form an $m$-dimensional vector space. We may assume that
$\langle v_{i}\vert v_{j}\rangle = \delta_{ij}$. Then $\left(v_1,\cdots,v_m
\right) \in \kansu{\stm}{\calh}$ and
\[
   F_0 \equiv \left\{\sum_{j=1}^{m}x_{j}v_{j}\vert x_j \in \fukuso \right\}
   \cong \fukuso^m.
\]
Namely, $F_0$ is a vector space associated with the o.n. basis
$\left(v_1,\cdots,v_m\right)$.

Next we assume for simplicity
that a family of unitary operators parameterized by $\calm$
\begin{equation}
   \label{eq:ufamily}
   W : \calm \rightarrow U(\calh),\quad W(\slam) = {\rm identity},
\end{equation}
connects $H_{\lam}$ and $H_0$  isospectrally:
\begin{equation}
  H_{\lam} \equiv W(\lam)H_0 W(\lam)^{-1}.
\end{equation}
In this case there is no level crossing of eigenvalues. Making use of
$W(\lam)$ we can define a projector
\begin{equation}
   \label{eq:pfamily}
  P : \calm \rightarrow \kansu{\grm}{\calh}, \quad
  P(\lam) \equiv
   W(\lam) \left(\sum^{m}_{j=1}v_{j}v_{j}^{\dagger}\right)W(\lam)^{-1}
\end{equation}
and  the pullback bundles over $\calm$:
\begin{equation}
   \label{eq:target}
  \left\{\kansu{U}{m},\widetilde{St}, \pi_{\widetilde{St}}, \calm\right\},\quad
  \left\{\fukuso^m,\widetilde{E}, \pi_{\widetilde{E}}, \calm\right\}.
\end{equation}

For the latter we set
\begin{equation}
   \label{eq:vacuum}
  \ket{vac} = \left(v_1,\cdots,v_m\right).
\end{equation}
In this case a canonical connection form $\cala$ of the principal bundle
$\left\{\kansu{U}{m},\widetilde{St}, \pi_{\widetilde{St}}, \calm\right\}$
is given by
\begin{equation}
   \label{eq:cform}
  \cala = \bra{vac}W(\lam)^{-1}d W(\lam)\ket{vac},
\end{equation}
where $d$ is a differential form on $\calm$
\[
d=\sum_{k}
\left(
d{\lam}_{k}\frac{\partial}{\partial \lam_{k}}+
d{{\bar \lam}}_{k}\frac{\partial}{\partial {\bar \lam}_{k}}
\right)
\]
together with its curvature form  (see \cite{SW} and \cite{MN})
\begin{equation}
   \label{eq:curvature-form}
   \calf \equiv d\cala+\cala\wedge\cala.
\end{equation}

Let $\gamma$ be a loop in $\calm$ at $\slam$, $\gamma : [0,1]
\rightarrow \calm, \gamma(0) = \gamma(1)=\slam$. For this $\gamma$ a holonomy
operator $\Gamma_{\cala}$ is defined:
\begin{equation}
   \label{eq:holonomy}
   \Gamma_{\cala}(\gamma) = {\cal P}exp\left\{\oint_{\gamma}\cala\right\}
   \in \kansu{U}{m},
\end{equation}
where ${\cal P}$ means path-ordering, see for example \cite{PZ}.
This acts on the fiber $F_0$ at $\slam$ of the vector bundle
$\left\{\fukuso^m,\widetilde{E}, \pi_{\widetilde{E}}, M\right\}$ as follows:
${\textbf x} \rightarrow \Gamma_{\cala}(\gamma){\textbf x}$.\quad
The holonomy group $Hol(\cala)$ is in general a subgroup of $\kansu{U}{m}$
. In the case of $Hol(\cala) = \kansu{U}{m}$,  $\cala$ is called
irreducible.
The irreducibility of $\cala$ is very important because it
means the universality of  quantum computation.
To check whether $\cala$ is irreducible or not we need its
curvature form (\ref{eq:curvature-form}), see \cite{MN}.

In the Holonomic Quantum Computation we take
\begin{eqnarray}
   \label{eq:information}
   &&{\rm Encoding\ of\ Information} \Longrightarrow {\textbf x} \in F_0 ,
   \nonumber \\
   &&{\rm Processing\ of\ Information} \Longrightarrow \Gamma_{\cala}(\gamma) :
   {\textbf x} \rightarrow \Gamma_{\cala}(\gamma){\textbf x}\equiv
   A{\textbf x}.
\end{eqnarray}
See the following figure.
%
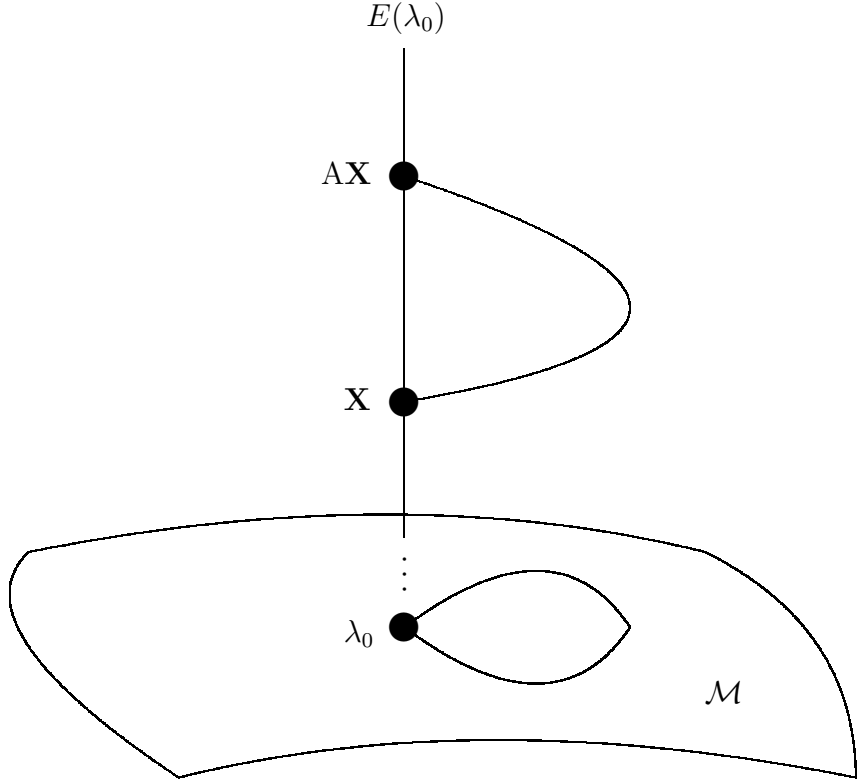
\begin{center}
\setlength{\unitlength}{1mm}
\begin{picture}(160,105)
\put(80,80){\circle*{4}}
\put(80,50){\circle*{4}}
\put(80,20){\circle*{4}}
\put(80,82){\line(0,1){15}}
\put(80,52){\line(0,1){36}}
\put(80,32){\line(0,1){26}}
\put(79.5,24){$\cdot$}
\put(79.5,26){$\cdot$}
\put(79.5,28){$\cdot$}
\qbezier(80,50)(140,60)(80,80)
\qbezier(80,20)(100,5)(110,20)
\qbezier(80,20)(100,35)(110,20)
\put(72,49){{\bf X}}
\put(69,79){A{\bf X}}
\put(72,18){$\lambda_{0}$}
\put(75,100){$E(\lambda_{0})$}
\qbezier(50,0)(90,10)(140,0)
\qbezier(50,0)(20,20)(30,30)
\qbezier(140,0)(140,20)(120,30)
\qbezier(30,30)(80,40)(120,30)
\put(120,10){${\cal M}$}
\end{picture}
\end{center}
%
%
Our model is relatively complicated compared to the other geometric models
and much more so than the usual spin models.
We have a lot of problems to solve in the near future.
We strongly hope young graduate students to invigorate this field.

\vspace{5mm}
\noindent
{\it Acknowledgment.}
The author wishes to thank Yoshinori Machida for his warm hospitality
at Numazu College of Technology. I also wishes to thank Akira Asada, 
Ryu Sasaki and Tatsuo Suzuki for reading this manuscript and making some 
useful comments.
%

\begin{center}
   \begin{Large}
      \textbf{Appendix}
   \end{Large}
\end{center}
\section*{A. A Family of Flag Manifolds}
\label{flag}
\setcounter{equation}{0}
\renewcommand{\theequation}{A.\arabic{equation}}

%
Let us make a comment on an interesting relation between flag manifolds
and the kernel of the exponential map defined on matrices. Here a (generalized)
flag manifold (which is a useful manifold as shown in the following)
is a natural generalization of the Grassmann one.

\par \noindent
First of all we make a brief review. For
\[
   \mbox{exp} : \real \longrightarrow S^{1}\ \subset \ \fukuso,  \quad
         \mbox{exp}(t)\equiv \mbox{e}^{2\pi \sqrt{-1}\, t}
\]
the kernel of this map is ker(exp)=$\integer \ \subset \ \real$.

We define by $H(n,\fukuso)$ the set of all hermitian matrices
\[
   H(n,\fukuso)=\{X\in M(n,\fukuso)\ |\ X^{\dagger}=X \}.
\]
Of course $H(1,\fukuso)=\real$. Note that each element of $H(n,\fukuso)$
can be diagonalized by some unitary matrix.

\par \noindent
The exponential map is now defined as
\begin{equation}
   \mbox{E} : H(n,\fukuso) \longrightarrow U(n) ,
       \quad \mbox{E}(X)=\mbox{e}^{2\pi  \sqrt{-1}X}.
\end{equation}
Here our target is ker(E).
\begin{flushleft}
{\bf Problem}\quad  What is the structure of ker(E) ?
\end{flushleft}
Our claim is that ker(E) is a family of flag manifolds. For that we write
ker(E) as
\begin{equation}
  K_{n}(\fukuso)=\{X \in H(n,\fukuso)\ |\ \mbox{e}^{2\pi  \sqrt{-1}X}={\bf 
1}_{n}\}.
\end{equation}
First we prove
\begin{equation}
   G_{n}(\fukuso)\ \subset \  K_{n}(\fukuso).
\end{equation}
Because since $P^{2}=P$ from the definition,  $P^{k}=P$ for $k \geq 1$,
so that
\begin{equation}
   \mbox{e}^{2\pi  \sqrt{-1}P} =
     {\bf 1}_{n}+ \sum_{k=1}^{\infty}\frac{(2\pi  \sqrt{-1})^k}{k!}P^{k}
   = {\bf 1}_{n}+ \sum_{k=1}^{\infty}\frac{(2\pi  \sqrt{-1})^k}{k!}P
   = {\bf 1}_{n}+(\mbox{e}^{2\pi  \sqrt{-1}}-1)P
   = {\bf 1}_{n}.
\end{equation}
We will prove that $G_{n}(\fukuso)$ becomes a kind of basis for
$K_{n}(\fukuso)$.

\par \noindent
For $X \in K_{n}(\fukuso)$ we write the set of all eigenvalues of $X$ as
$\mbox{spec}(X)$. Then $\mbox{spec}(X)$ = $\{0, 1\}$ for
$X \in G_{n}(\fukuso)$.
\par \noindent
It is clear that $\mbox{spec}(X) \subset \integer$. For $X \in K_{n}(\fukuso)$
we have
\begin{equation}
    \mbox{spec}(X)=\{n_1(d_1),\ \cdots ,\ n_k(d_k),\ \cdots , n_j(d_j)\}\quad
  \mbox{where}\quad
       n_k \in \integer\ \mbox{and}\ \sum_{k=1}^{j}d_{k}=n,
\label{specX}
\end{equation}
in which ($d_k$) is the multiplicity of the eigenvalue $n_k$.
Since $X$ is diagonalized by some $U \in U(n)$,
\begin{equation}
X = U
     X_0
     U^{-1}
    = \sum_{k=1}^{j}n_kP_{d_k},
\end{equation}
where
\begin{equation}
X_{0} =
     \left(
       \begin{array}{ccccc}
        n_1{\bf 1}_{d_1}&     &      &  &    \\
          &n_2{\bf 1}_{d_2}&      &  &     \\
          &     &\cdot &  &                \\
          &     &   &\quad \cdot&               \\
          &     &      &   &\quad n_j{\bf 1}_{d_j}  \\
       \end{array}
      \right),\quad
  P_{d_k}= U
     \left(
       \begin{array}{ccccc}
          {\bf 0}_{d_1}&     &   &  &     \\
          &\cdot & & &                    \\
          & &{\bf 1}_{d_k}&   &           \\
          &     & & \cdot &               \\
          &     &      &    &{\bf 0}_{d_j}
       \end{array}
      \right)
     U^{-1}.
\end{equation}

\vspace{3mm} \par \noindent
Here we list some properties of the set of projections $\{P_{d_k}\}$:
\[
   \mbox{(1)}\quad  P_{d_k} \in G_{d_k,n}(\fukuso),\quad
   \mbox{(2)}\quad  P_{d_k}P_{d_l}=\delta_{kl}P_{d_l},\quad
   \mbox{(3)}\quad  P_{d_1}+ P_{d_2}+ \cdots + P_{d_j}= {\bf 1}_{n}.
\]
%
%
Let us here prepare a terminology. For $X \in K_{n}(\fukuso)$ we call set 
of the
eigenvalues together with  multiplicities
\[
      \{(n_1,d_1),(n_2,d_2), \cdots , (n_j,d_j)\}
\]
the spectral type of $X$.

\par \noindent
Then it is easy to see that
$X$ and $Y$ $\in K_{n}(\fukuso)$ are of the same spectral type ($X \sim Y$)
if and only if $Y=UXU^{-1}$ for some $U \in U(n)$.
For $X \in K_{n}(\fukuso)$ we define
\[
   C(X)=\{Y \in K_{n}(\fukuso)\ |\  Y \sim X \}\ .
\]
We have clearly $C(X)=C(X_0)$. Then it is easy to see that
$K_{n}(\fukuso)$ can be classified by the spectral type
\begin{equation}
  \label{eq:equivalent class}
   K_{n}(\fukuso)=\bigcup_{X}C(X)=\bigcup_{X_0}C(X_0)
\end{equation}
and the unitary group $U(n)$ acts on $C(X)$ as follows:
\[
    U(n)\times C(X) \longrightarrow C(X)\ :\  (U,X)\mapsto UXU^{-1}.
\]
Since this action is free and transitive, the isotropy group at $X_0$ is
\[
     U(d_1)\times U(d_2)\times \cdots \times U(d_j)\ ,
\]
so that we have
\begin{equation}
      C(X)\cong \frac{U(n)}{U(d_1)\times U(d_2)\times \cdots \times U(d_j)}\ .
\end{equation}
The right hand side is called a generalized flag manifold. In particular
when $d_1=d_2=\cdots=d_n=1$ (there is no overlapping in the eigenvalues of
X) we have
\begin{equation}
      C(X)\cong \frac{U(n)}{U(1)\times U(1)\times \cdots \times U(1)}\ .
\end{equation}
This is called a flag manifold.

\par \noindent
Namely by (\ref{eq:equivalent class}) we know that $K_{n}(\fukuso)$ is a
family of generalized flag manifolds.

A comment is in order.
For the Grassmann manifolds we have very good local coordinates like
(\ref{eq:local-coordinate}), while we don't know  good local
coordinates for generalized flag manifolds.
\begin{flushleft}
{\bf Problem}\quad  Find a good local coordinate system.
\end{flushleft}
For some applications of generalized flag manifolds
the paper \cite{RFP} is recommended. See also  \cite{RFP} and  references 
therein.

\section*{B. A Generalization of Pauli Matrices}
\label{pauli}
\setcounter{equation}{0}
\renewcommand{\theequation}{B.\arabic{equation}}

%
Here let us introduce a generalization of Pauli matrices
(\ref{eq:pauli-matrices}) which has been used in several situations in
both Quantum Field Theory and Quantum Computation.

First of all we summarize the properties of Pauli matrices. By (\ref
{eq:pauli-matrices})
$\sigma_{2}=\sqrt{-1}\sigma_{1}\sigma_{3}$, so that the essential elements
of Pauli matrices are $\{\sigma_{1}, \sigma_{3}\}$ and they satisfy
\begin{equation}
\sigma_{1}^{2}=\sigma_{3}^{2}={\bf 1}_{2};\quad  \quad
\sigma_{1}^{\dagger}=\sigma_{1},\
\sigma_{3}^{\dagger}=\sigma_{3};\quad \quad
\sigma_{3}\sigma_{1}=- \sigma_{1}\sigma_{3}.
\end{equation}

\par \noindent
Let $\{\Sigma_{1}, \Sigma_{3}\}$ be the following matrices in $M(n,\fukuso)$
\begin{equation}
\Sigma_{1}=
\left(
\begin{array}{ccccccc}
0&  &  &      &      &      & 1  \\
1& 0&  &      &      &      &    \\
  & 1& 0&      &      &      &    \\
  &  & 1& \cdot&      &      &    \\
  &  &  & \cdot& \cdot&      &    \\
  &  &  &      & \cdot& \cdot&    \\
  &  &  &      &      &    1 & 0
\end{array}
\right),      \qquad
\Sigma_{3}=
\left(
\begin{array}{ccccccc}
1&       &           &      &      &      &             \\
  & \sigma&           &      &      &      &             \\
  &       & {\sigma}^2&      &      &      &             \\
  &       &           & \cdot&      &      &             \\
  &       &           &      & \cdot&      &             \\
  &       &           &      &      & \cdot&             \\
  &       &           &      &      &      & {\sigma}^{n-1}
\end{array}
\right)
\end{equation}
where $\sigma$ is a primitive root of unity ${\sigma}^{n}=1$ (
$\sigma=\mbox{e}^{\frac{2\pi \sqrt{-1}}{n}}$). We note that
\[
\bar{\sigma}=\sigma^{n-1},\quad
1+\sigma+\cdots+\sigma^{n-1}=0 .
\]
The two matrices
$\{\Sigma_{1}, \Sigma_{3}\}$ are generalizations of Pauli matrices
$\{\sigma_{1}, \sigma_{3}\}$, but they are not hermitian.
Here we list some of their important properties:
\begin{equation}
\Sigma_{1}^{n}=\Sigma_{3}^{n}={\bf 1}_{n}\quad ; \quad
\Sigma_{1}^{\dagger}=\Sigma_{1}^{n-1},\
\Sigma_{3}^{\dagger}=\Sigma_{3}^{n-1}\quad ; \quad
\Sigma_{3}\Sigma_{1}=\sigma \Sigma_{1}\Sigma_{3}\ .
\end{equation}
If we define a Vandermonde matrix $W$ based on $\sigma$ as
\begin{eqnarray}
\label{eq:Large-double}
W&=&\frac{1}{\sqrt{n}}
\left(
\begin{array}{ccccccc}
1&        1&     1&   \cdot & \cdot  & \cdot & 1             \\
1& \sigma^{n-1}& \sigma^{2(n-1)}&  \cdot& \cdot& \cdot& \sigma^{(n-1)^2} \\
1& \sigma^{n-2}& \sigma^{2(n-2)}&  \cdot& \cdot& \cdot& \sigma^{(n-1)(n-2)} \\
\cdot&  \cdot &  \cdot  &     &      &      & \cdot  \\
\cdot&  \cdot  & \cdot &      &      &      &  \cdot  \\
1& \sigma^{2}& \sigma^{4}& \cdot & \cdot & \cdot & \sigma^{2(n-1)} \\
1& \sigma & \sigma^{2}& \cdot& \cdot& \cdot& \sigma^{n-1}
\end{array}
\right), \\
\label{eq:Large-double-dagger}
W^{\dagger}&=&\frac{1}{\sqrt{n}}
\left(
\begin{array}{ccccccc}
1&        1&     1&   \cdot & \cdot  & \cdot & 1             \\
1& \sigma& \sigma^{2}&  \cdot& \cdot& \cdot& \sigma^{n-1} \\
1& \sigma^{2}& \sigma^{4}&  \cdot& \cdot& \cdot& \sigma^{2(n-1)} \\
\cdot&  \cdot &  \cdot  &     &      &      & \cdot  \\
\cdot&  \cdot  & \cdot &      &      &      &  \cdot  \\
1& \sigma^{n-2}& \sigma^{2(n-2)}& \cdot& \cdot& \cdot& \sigma^{(n-1)(n-2)} \\
1&    \sigma^{n-1} & \sigma^{2(n-1)}& \cdot& \cdot& \cdot& \sigma^{(n-1)^2}
\end{array}
\right),
\end{eqnarray}
then it is not difficult to see
\begin{equation}
\label{eq:diagonalization}
\Sigma_{1}=W\Sigma_{3}W^{\dagger}=W\Sigma_{3}W^{-1}.
\end{equation}
For example, for $n=3$
\begin{eqnarray}
W\Sigma_{3}W^{\dagger}&=&\frac{1}{3}
\left(
\begin{array}{ccc}
1&          1&     1     \\
1& \sigma^{2}& \sigma    \\
1& \sigma    & \sigma^{2}
\end{array}
\right)
\left(
\begin{array}{ccc}
1&       &           \\
  & \sigma&           \\
  &       & {\sigma}^2
\end{array}
\right)
\left(
\begin{array}{ccc}
1&        1&     1 \\
1& \sigma & \sigma^{2} \\
1& \sigma^{2}  & \sigma
\end{array}
\right)   \nonumber \\
&=&\frac{1}{3}
\left(
\begin{array}{ccc}
1& \sigma& \sigma^{2}\\
1&      1&  1        \\
1& \sigma^{2}& \sigma
\end{array}
\right)
\left(
\begin{array}{ccc}
1&        1&     1 \\
1& \sigma& \sigma^{2} \\
1& \sigma^{2}  & \sigma
\end{array}
\right)
=\frac{1}{3}
\left(
\begin{array}{ccc}
0&  0& 3\\
3&  0& 0\\
0&  3& 0
\end{array}
\right)
=\Sigma_{1},  \nonumber
\end{eqnarray}
where we have used that $\sigma^{3}=1,\ \bar{\sigma}=\sigma^{2}\ \mbox{and}\
1+\sigma+\sigma^{2}=0$.

\par \noindent
That is, $\Sigma_{1}$ can be diagonalized by making use of $W$.

\par \noindent
A comment is in order. Since $W$ corresponds to the Walsh-Hadamard matrix
(\ref{eq:w-a}), so it may be possible to call $W$
the generalized Walsh-Hadamard matrix.

\section*{C.  General Controlled Unitary Operations}
\label{gcuo}
\setcounter{equation}{0}
\renewcommand{\theequation}{C.\arabic{equation}}

%
%
Here let us introduce a usual construction of general controlled unitary
operations to help the understanding of general controlled NOT one.
In the following arguments if we take $U=X=\sigma_{1}$ then they reduce to
the arguments of a construction of general controlled NOT operator.

First of all let us recall (\ref{eq:mod2-identity}).
For $x, y, z \in \integer_{2}$ we have identities :
\begin{eqnarray}
\label{eq:mod2-identity-1}
  &&x + y - x\oplus y = 2xy, \\
\label{eq:mod2-identity-2}
  &&x + y + z - x\oplus y - x\oplus z - y\oplus z + x\oplus y \oplus z = 4xyz,
\end{eqnarray}
where $x\oplus y=x+y\ (\mbox{mod}\ 2)$.
For the most general identities of above type see \cite{9th} and \cite{KF6}.

\par \noindent
The controlled-controlled unitary operations are constructed by making use
of both several controlled unitary operations and controlled NOT operations:
Let $U$ be an arbitrarily unitary matrix in $U(2)$ and $V$ a unitary one
in $U(2)$ satisfying $V^2=U$.  Then by (\ref{eq:mod2-identity-1}) we have
\begin{eqnarray}
V^{x+y-x\oplus y}&=&V^{2xy}=(V^2)^{xy}=U^{xy},  \\
V^{x+y-x\oplus y}&=&V^{x}V^{y}V^{-x\oplus y}
                   =V^{x}V^{y}(V^{-1})^{x\oplus y}
                   =V^{x}V^{y}(V^{\dagger})^{x\oplus y},
\end{eqnarray}
so a controlled-controlled $U$ operation is graphically represented as
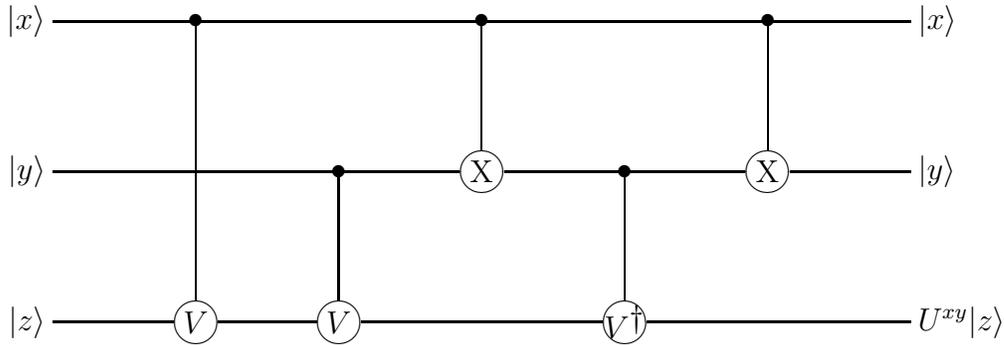
\begin{center}
\setlength{\unitlength}{1mm}
\begin{picture}(150,55)
\put(10,50){\line(1,0){114}}   
\put(10,30){\line(1,0){54}}   
\put(70,30){\line(1,0){32}}   
\put(108,30){\line(1,0){16}}   
\put(10,10){\line(1,0){16}}   
\put(32,10){\line(1,0){13}}   
\put(51,10){\line(1,0){32}}   
\put(89,10){\line(1,0){35}}   
\put(0,45){\makebox(9,10)[r]{$|x\rangle$}} 
\put(0,25){\makebox(9,10)[r]{$|y\rangle$}} 
\put(0,5){\makebox(9,10)[r]{$|z\rangle$}} 
\put(125,45){\makebox(25,10)[l]{$|x\rangle$}} 
\put(125,25){\makebox(25,10)[l]{$|y\rangle$}} 
\put(125,5){\makebox(25,10)[l]{$U^{xy}|z\rangle$}}
\put(29,13){\line(0,1){37}}     
\put(48,13){\line(0,1){17}}     
\put(67,33){\line(0,1){17}}     
\put(86,13){\line(0,1){17}}     
\put(105,33){\line(0,1){17}}     
\put(26,45){\makebox(6,10){$\bullet$}} 
\put(45,25){\makebox(6,10){$\bullet$}} 
\put(64,45){\makebox(6,10){$\bullet$}} 
\put(83,25){\makebox(6,10){$\bullet$}} 
\put(102,45){\makebox(6,10){$\bullet$}} 
\put(29,10){\circle{6}}               
\put(48,10){\circle{6}}               
\put(67,30){\circle{6}}               
\put(86,10){\circle{6}}               
\put(105,30){\circle{6}}               
\put(64,25){\makebox(6,10){X}}         
\put(102,25){\makebox(6,10){X}}         
\put(26,5){\makebox(6,10){$V$}}         
\put(45,5){\makebox(6,10){$V$}}         
\put(83,5){\makebox(6,10){$V^{\mbox{\dag}}$}}   
\end{picture}
\end{center}
You should read this figure as follows: From the left to the right
\[
\ket{x}\otimes \ket{y}\otimes \ket{z}\ \longrightarrow \
\ket{x}\otimes \ket{y}\otimes U^{xy}\ket{z}.
\]

\par \noindent
The controlled-controlled-controlled unitary operations are constructed
as follows:
Let $U$ be an arbitrarily unitary matrix in $U(2)$ and $V$ be a unitary
one in $U(2)$ satisfying $V^4=U$. Then by (\ref{eq:mod2-identity-1})
\begin{eqnarray}
V^{x+y+z-(x\oplus y+x\oplus z+y\oplus z)+x\oplus y\oplus z}&=&
V^{4xyx}=U^{xyz}, \\
V^{x+y+z-(x\oplus y+x\oplus z+y\oplus z)+x\oplus y\oplus z}&=&
V^{x}V^{y}V^{z}(V^{\dagger})^{x\oplus y}(V^{\dagger})^{x\oplus z}
(V^{\dagger})^{y\oplus z}V^{x\oplus y\oplus z},
\end{eqnarray}
so a controlled-controlled-controlled $U$ operation is graphically
represented as
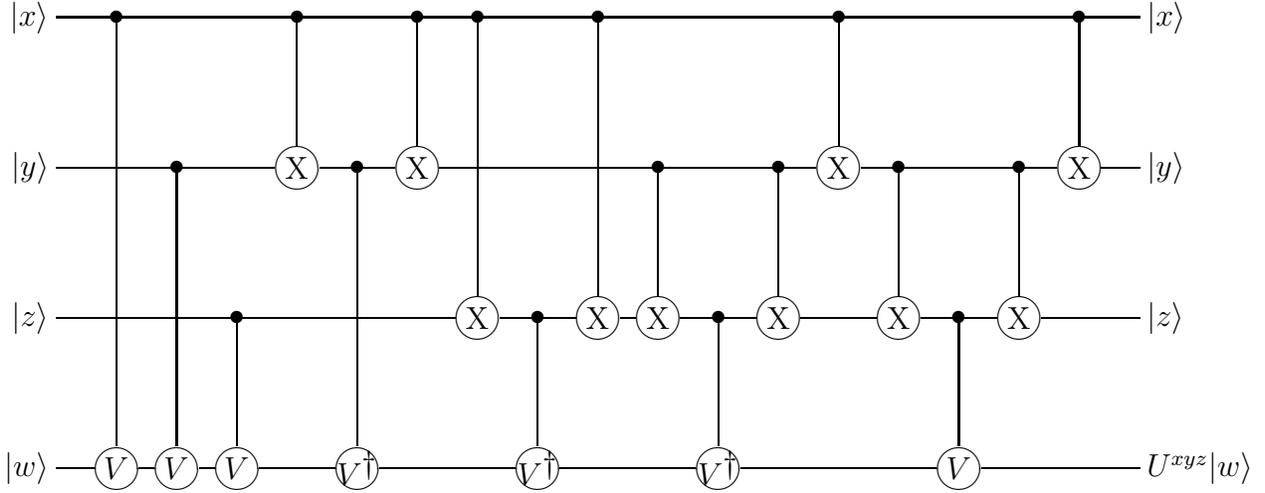
\begin{center}
\setlength{\unitlength}{1mm}
\begin{picture}(160,80)
\put(6,70){\line(1,0){144}}   
\put(6,50){\line(1,0){29}}   
\put(41,50){\line(1,0){10}}   
\put(57,50){\line(1,0){50}}   
\put(113,50){\line(1,0){26}}   
\put(145,50){\line(1,0){5}}   
\put(6,30){\line(1,0){53}}   
\put(65,30){\line(1,0){10}}   
\put(81,30){\line(1,0){2}}   
\put(89,30){\line(1,0){10}}   
\put(105,30){\line(1,0){10}}   
\put(121,30){\line(1,0){10}}   
\put(137,30){\line(1,0){13}}   
\put(6,10){\line(1,0){5}}   
\put(17,10){\line(1,0){2}}   
\put(25,10){\line(1,0){2}}   
\put(33,10){\line(1,0){10}}   
\put(49,10){\line(1,0){18}}   
\put(73,10){\line(1,0){18}}   
\put(97,10){\line(1,0){26}}   
\put(129,10){\line(1,0){21}}   
\put(0,65){\makebox(5,10)[r]{$|x\rangle$}} 
\put(0,45){\makebox(5,10)[r]{$|y\rangle$}} 
\put(0,25){\makebox(5,10)[r]{$|z\rangle$}} 
\put(0,5){\makebox(5,10)[r]{$|w\rangle$}} 
\put(151,65){\makebox(9,10)[l]{$|x\rangle$}} 
\put(151,45){\makebox(9,10)[l]{$|y\rangle$}} 
\put(151,25){\makebox(9,10)[l]{$|z\rangle$}} 
\put(151,5){\makebox(9,10)[l]{$U^{xyz}|w\rangle$}}
\put(14,13){\line(0,1){57}}     
\put(22,13){\line(0,1){37}}     
\put(30,13){\line(0,1){17}}     
\put(38,53){\line(0,1){17}}     
\put(46,13){\line(0,1){37}}     
\put(54,53){\line(0,1){17}}     
\put(62,33){\line(0,1){37}}     
\put(70,13){\line(0,1){17}}     
\put(78,33){\line(0,1){37}}     
\put(86,33){\line(0,1){17}}     
\put(94,13){\line(0,1){17}}     
\put(102,33){\line(0,1){17}}     
\put(110,53){\line(0,1){17}}     
\put(118,33){\line(0,1){17}}     
\put(126,13){\line(0,1){17}}     
\put(134,33){\line(0,1){17}}     
\put(142,53){\line(0,1){17}}     
\put(11,65){\makebox(6,10){$\bullet$}} 
\put(19,45){\makebox(6,10){$\bullet$}} 
\put(27,25){\makebox(6,10){$\bullet$}} 
\put(35,65){\makebox(6,10){$\bullet$}} 
\put(43,45){\makebox(6,10){$\bullet$}} 
\put(51,65){\makebox(6,10){$\bullet$}} 
\put(59,65){\makebox(6,10){$\bullet$}} 
\put(67,25){\makebox(6,10){$\bullet$}} 
\put(75,65){\makebox(6,10){$\bullet$}} 
\put(83,45){\makebox(6,10){$\bullet$}} 
\put(91,25){\makebox(6,10){$\bullet$}} 
\put(99,45){\makebox(6,10){$\bullet$}} 
\put(107,65){\makebox(6,10){$\bullet$}} 
\put(115,45){\makebox(6,10){$\bullet$}} 
\put(123,25){\makebox(6,10){$\bullet$}} 
\put(131,45){\makebox(6,10){$\bullet$}} 
\put(139,65){\makebox(6,10){$\bullet$}} 
\put(14,10){\circle{6}}               
\put(22,10){\circle{6}}               
\put(30,10){\circle{6}}               
\put(38,50){\circle{6}}               
\put(46,10){\circle{6}}               
\put(54,50){\circle{6}}               
\put(62,30){\circle{6}}               
\put(70,10){\circle{6}}               
\put(78,30){\circle{6}}               
\put(86,30){\circle{6}}               
\put(94,10){\circle{6}}               
\put(102,30){\circle{6}}               
\put(110,50){\circle{6}}               
\put(118,30){\circle{6}}               
\put(126,10){\circle{6}}               
\put(134,30){\circle{6}}               
\put(142,50){\circle{6}}               
\put(35,45){\makebox(6,10){X}}         
\put(51,45){\makebox(6,10){X}}         
\put(107,45){\makebox(6,10){X}}         
\put(139,45){\makebox(6,10){X}}         
\put(59,25){\makebox(6,10){X}}         
\put(75,25){\makebox(6,10){X}}         
\put(83,25){\makebox(6,10){X}}         
\put(99,25){\makebox(6,10){X}}         
\put(115,25){\makebox(6,10){X}}         
\put(131,25){\makebox(6,10){X}}         
\put(11,5){\makebox(6,10){$V$}}         
\put(19,5){\makebox(6,10){$V$}}         
\put(27,5){\makebox(6,10){$V$}}         
\put(43,5){\makebox(6,10){$V^{\mbox{\dag}}$}}         
\put(67,5){\makebox(6,10){$V^{\mbox{\dag}}$}}         
\put(91,5){\makebox(6,10){$V^{\mbox{\dag}}$}}         
\put(123,5){\makebox(6,10){$V$}}        
\end{picture}
\end{center}
This figure means
\[
\ket{x}\otimes \ket{y}\otimes \ket{z}\otimes \ket{w}
\ \longrightarrow \
\ket{x}\otimes \ket{y}\otimes \ket{z}\otimes U^{xyz}\ket{w}.
\]

\par \vspace{3mm} \noindent
A comment is in order. For the case $U=X=\sigma_{1}$ we have from
(\ref{eq:repeated-controlled-not})
\begin{eqnarray}
\ket{a_1,a_2,\cdots,a_{n-1},a_{n}}\ \longrightarrow
&&\ket{a_1,a_2,\cdots,a_{n-1},a_{1}a_{2}\cdots a_{n-1}\oplus a_n} \nonumber \\
\equiv
&&\ket{a_{1}}\otimes \ket{a_{2}}\otimes \cdots \ket{a_{n-1}}\otimes
\ket{a_{1}a_{2}\cdots a_{n-1}\oplus a_n}
\nonumber \\
=
&&\ket{a_{1}}\otimes \ket{a_{2}}\otimes \cdots \ket{a_{n-1}}\otimes
X^{a_{1}a_{2}\cdots a_{n-1}}\ket{a_{n}}.
\end{eqnarray}

As can be seen from the figures the well-known construction of general
controlled unitary operations needs exponential steps. Namely it is not
efficient. For more details see \cite{9th} and \cite{KF6}.


\end{document}